\documentclass[twocolumn]{aastex701}
\usepackage{hyperref}
\usepackage{amsmath}
\usepackage{float}
\usepackage{algorithm}
\usepackage{algpseudocode}

\newcommand{\lcdm}{$\Lambda$CDM}
\newcommand{\hmpc}{$h^{-1}\,$Mpc}

\newcommand{\vide}{\textsc{vide}}
\newcommand{\videone}{\textsc{vide 1.0}}
\newcommand{\videtwo}{\textsc{vide 2.0}}
\newcommand{\healpix}{\textsc{HEALPix}}
\newcommand{\nside}{\textsc{nside}} 
\newcommand{\bigO}{\mathcal{O}}
\newcommand{\ap}{Alcock--Paczy\'nski}

\begin{document}

\title{VIDE 2.0: Robust, Scalable Void Finding for Extremely Large Galaxy Surveys}

\author[gname='Paul',sname='Sutter']{Paul M. Sutter}
\affiliation{Department of Physics and Astronomy, Johns Hopkins University, Baltimore, MD 21218, USA}
\email[show]{psutter1@jhu.edu}

\author[gname='Benjamin',sname='Wandelt']{Benjamin D. Wandelt}
\affiliation{Department of Physics and Astronomy, Johns Hopkins University, Baltimore, MD 21218, USA}
\affiliation{Department of Applied Mathematics and Statistics, Johns Hopkins University, Baltimore, MD 21218, USA}
\email[show]{wandelt@jhu.edu}

\begin{abstract}

Current and future spectroscopic and photometric galaxy surveys will deliver
unprecedented numbers of galaxies, with expected counts in the tens of
millions. To answer the computational challenges posed by these surveys, in
this paper we present \videtwo, a significantly improved and enhanced void
finding algorithm and analysis pipeline based on \vide~\citep{Sutter2014}. Leveraging the
parameter-free identification of voids via a Voronoi tessellation and watershed
scheme, we implement a robust mechanism for handling survey boundaries and
incompleteness without relying on vast numbers of boundary particles. This
new method identifies edge galaxies and flags voids contaminated by their
presence. The mechanism is fully deterministic, producing bit-for-bit
reproducible catalogs from identical inputs. An automatic
rule for setting the mask resolution renders the entire pipeline
parameter-free. Combined with other
performance enhancements to the underlying \vide~engine,
we find significant ($> 100\times$) speedups
and a dramatic reduction in memory usage compared to previous versions: a
12-million-galaxy DESI~Y1-class mock catalog now completes in roughly 15
minutes on a single laptop, where \videone~could not run a catalog of even
$10^6$ galaxies on the same hardware. We are also able to identify voids
existing closer to survey boundaries via deterministic per-void contamination
flags, maximizing their statistical potential. We exercise the upgraded
pipeline on two large mock suites (MultiDark-Patchy DR12 and AbacusSummit
DESI~Y1 ELG), demonstrating converged cosmic-variance covariances for the
void number function, the void-galaxy cross-correlation, and the void
ellipticity distribution --- key ingredients for Stage~IV void
cosmology. \videtwo~is now well suited to handle large
numbers of galaxies in mock and real observations, and ready to tackle the
challenges of future campaigns. The code described in this paper is
available at \url{www.cosmicvoids.net}.

\end{abstract}


\section{Introduction}
\label{sec:intro}

The study of cosmic voids, the vast underdense regions that dominate the volume of the
universe~\citep{Thompson2011}, has grown into a
broad field encompassing many robust cosmological probes.
Owing to their relatively simple evolution and pristine
contents~\citep{Sutter2014b,Schuster2025,Wang2024}, voids
provide complementary constraints on cosmological parameter
space~\citep{Bromley2025}, a window into the evolution of low-mass
galaxies~\citep{Shojaei2025}, an opportunity to independently constrain
neutrino mass~\citep{Massara_2015,Verza2023,Bayer2024}, a pathway to testing gravity free
from cluster systematics~\citep{Wilson2021,Wilson2023,Williams2025}, as well as
a laboratory for investigating galaxy bias~\citep{Montero-Dorta2025} and
alignment~\citep{DAssigniesD2022}, among many other strengths. Besides raw number
counts~\citep{Pisani2015,Kreisch2022}, many void properties are rich with
information, such as their density
profiles~\citep{Hamaus2014,Sutter2014c,Curtis2025}, the \ap\ test~\citep{Lavaux_2012}, redshift space distortions~\citep{Hamaus_2017}, spin
distributions~\citep{Kang2025}, and cross-correlations with gravitational lensing
shear~\citep{Hossen2022,Maggiore2025} and gamma-ray
sources~\citep{Furniss2025}.

In~\citet{Sutter2012} we introduced a publicly available void catalog, drawn
from the SDSS DR7 main sample~\citep{Abazajian2009}. Since then, additional
public catalogs~(e.g., \cite{Mao2016,Douglass2023}) have continued
to enrich and enliven the void ecosystem. These catalogs have been analyzed in multiple subsequent studies, for example for cosmological constraints from
measurements of the \ap~effect~\citep{Sutter2014,Hamaus2016} and the void size
distributions~\citep{Contarini2023,Song2025}, with galaxy-void cross-correlations leading to precision constraints on the expansion geometry and growth of structure~\citep{2020JCAP...12..023H}.

The first data release of DESI (DR1)~\citep{DESICollaboration2025} has inaugurated the era of Stage IV cosmological surveys, which will include observing campaigns with the Euclid space telescope~\citep{Mellier2025}, the SPHEREx mission~\citep{2026ApJ...999..139B}, the Nancy Grace Roman Space Telescope~\citep{Schlieder2024}, and the Chinese Space Station Survey~\citep{Gong2025}. These massive surveys will unlock unprecedented numbers of galaxies, typically tens of millions per survey. Mock observations and theoretical modeling suggest that voids in these surveys will provide extraordinary constraints through void number functions~\citep{Contarini2022, Verza2025}, the void power spectrum~\citep{Song2024}, the void-lensing cross-correlation~\citep{Xiong2025}, the void-galaxy cross-correlation~\citep{Hamaus_2014}, and more. 

To achieve these predicted results, void finders must provide reliable,
well-understood void populations and statistics that allow robust tracing from
void properties identified in galaxy surveys to the underlying dark matter
distribution, and further to the fundamental cosmological parameters.
Additionally, any void finder must grapple with the large volumes of data
presented by existing and upcoming surveys while dealing with the vagaries of
real-world data collection, including boundary mask handling and survey
incompleteness.

Historically, void finders were developed for dark matter simulations (or halo
or galaxy populations drawn from the dark matter), which provided tracer
densities to arbitrary resolution while side-stepping observational
effects (e.g., \cite{Neyrinck2008}). Meanwhile, pioneering efforts
identified voids observationally with simple approximations from limited survey
data~(e.g., \cite{Kauffmann1991,2002ApJ...566..641H}). In~\citet{Sutter2014} we introduced~\vide, the
Void IDentification and Examination toolkit, a public void finder based on a
heavily modified version of the \textsc{zobov} algorithm employed
in~\citet{Neyrinck2008} that could handle realistic survey scenarios. This
engine, which discovered voids via a parameter-free tessellation scheme to
build a density field from the given tracers, followed by a watershed transform
to join adjacent regions into void basins, powered a raft of void discoveries
in both simulations and observations, including many of the above references.

In addition to performance and reliability improvements, alongside extensions to handle arbitrary volumes and sparse density sampling, a key innovation behind \vide~compared to \textsc{zobov} was its handling of survey boundaries, which
involved flooding edges with vast numbers of boundary particles to constrain
the tessellation. This approach preserved (as much as possible) the
parameter-free spirit of void analysis, and it became
the \emph{de facto} method employed to produce other public
catalogs (e.g., \cite{Mao2016}).

While this approach worked well for limited galaxy surveys, for large (e.g., Stage IV) surveys, this approach to boundary handling produces far more boundary particles than actual galaxies, requiring drastically more memory and compute for a typical tessellation; enough to lead to impractically long run-times, if not outright overwhelming typical system constraints.

This procedure introduced an element of stochasticity, as the algorithm randomly placed the boundary particles, which would lead to small catalog variations from run to run. 

It also required an extra tunable parameter, the density
of boundary particles, that propagated into distortions of the tessellation and
thereby the resulting void catalog. Meanwhile, other void finders suitable for surveys have emerged, including approaches leveraging persistent
homology~\citep{Aggarwal2024}, dynamical modeling~\citep{Monllor-Berbegal2025},
spherical aggregation~\citep{Rincon2025} and graph flows~\citep{2026OJAp....960855T}. Several studies have compared
and contrasted these differing
algorithms~(e.g., \cite{Colberg2008,Veyrat2023,Douglass2023,Zaidouni2025}),
which generally find broad agreement of key void properties (e.g., the location
of density minima). However, some ambiguities persist, which must be resolved to
fully unlock the potential of cosmic voids.

To answer the data volume challenges presented by modern and future surveys,
and to provide a robust, flexible platform for analysis and comparison, in this
paper we present~\videtwo, an upgraded and enhanced implementation of our original
void-finding algorithm. This upgrade includes several performance
enhancements, a new tessellation engine, and most crucially, a 
parameter-free, deterministic approach to boundary handling that preserves the scale-agnostic, purely morphological void definition in any survey configuration.

This paper is organized as follows. In Section~\ref{sec:masks} we detail the
new boundary handling procedure. In Section~\ref{sec:scaling} we showcase the
improved scalability and performance of the new scheme and the entire
\vide~pipeline. In Section~\ref{sec:impacts} we elucidate how the new boundary
handling approach impacts the underlying tessellation and resulting void
catalogs (in particular emphasizing how the new void catalogs are more
reliable). In Section~\ref{sec:demos} we run the upgraded pipeline
on two large mock suites and measure the void statistics and
covariances that Stage~IV void cosmology will require. We conclude in Section~\ref{sec:conclusions}
by providing general philosophies for approaching void catalogs and
summarize forward-looking prospects and future work.

\section{Parameter Free Survey Boundary Handling}
\label{sec:masks}

\subsection{The original procedure and its limitations}
\label{sec:masks_v1}

The primary strategy behind \vide's approach to void finding is to treat them
as objects that are defined in a scale-agnostic way, only by the morphology of the tracer distribution. We implement this philosophy through a heavily modified
version of \textsc{zobov}~\citep{Neyrinck2008}, which begins with a tessellation
step to transform a set of tracers (e.g., dark matter particles in a simulation
or galaxies in a survey) into a density field. We then apply the watershed
transform to that density field to segment it into a series of basins (which
become the voids) separated by ridges (the void boundaries).

This tracer-agnostic approach offers many advantages, which are elucidated in
more detail in~\citet{Sutter2014}. Primarily, \vide~is parameter-free. There is
a single, \emph{optional} tunable parameter that controls the threshold by
which adjacent basins are merged into singular voids. In practice, adjusting
this value has little impact, especially in galaxy surveys where tracer density
is relatively sparse, as \vide~naturally produces a hierarchy of voids and
sub-voids; adjusting the merging threshold merely changes the truncation point
for the identification of low-contrast super-voids spanning large volumes.

As is usually the case, to make this ideal philosophy useful in realistic
scenarios, such as in a galaxy catalog that results from a real survey, we must contend with the limitations of actual observational data. In particular, the presence of a survey mask alters the topology of
the density field. It introduces artificial bounds, beyond which the density
field is unknown, spatially varying radial survey completeness, and foreground objects insert punctures that must be accounted for.

Since we understand voids as basins in the density field, these
effects necessarily alter the sizes, shapes, and distributions of
identified voids~\citep{Sutter2014d}. For example, masks preferentially
eliminate large voids while preserving smaller ones. They also introduce a more
subtle bias, as voids that happen to lie along the mask edge survive intact,
while those perpendicular to the mask become truncated. This can impact
analyses that require a fair sampling of void orientations (e.g., the
\ap~effect).

To be fair, the same problems affect other void finding algorithms, since
regardless of underlying philosophy voids are large, extended objects.

When we originally adapted \vide~for galaxy surveys, we developed a boundary
handling scheme that was primarily designed to be as
conservative and cautious as possible. The scheme was as follows:

\begin{enumerate}
  \item Convert a given survey mask to \healpix~format with a chosen \nside~resolution.
  \item Identify the boundaries of that mask. This step is illustrated in Figure~\ref{fig:mask}, and will be identical for our new implementation.
  \item ``Flood the zone'' by placing \textit{boundary particles}, uniformly at random with a given density along the radial rays defined by boundary pixel centers.
  \item Also place boundary particles uniformly at random along the maximum and minimum redshift caps of the survey volume.
  \item Proceed with the Voronoi tessellation, including the inserted boundary particles.
  \item Prior to the watershed step, remove any adjacency links to any galaxy that is itself adjacent to a boundary particle (i.e., in the local Voronoi tessellation sense), effectively preventing the voids from reaching mask edges. 
\end{enumerate}

In practice, this procedure was destructive: any galaxy adjacent to a
boundary particle had its Voronoi volume set to effectively zero and all
adjacency links severed, permanently erasing information from
the density field. The resulting void catalog was thus shaped not only by
the real galaxy distribution, but also by the particular random
realization of boundary particles.

\begin{figure}[htbp]
    \plotone{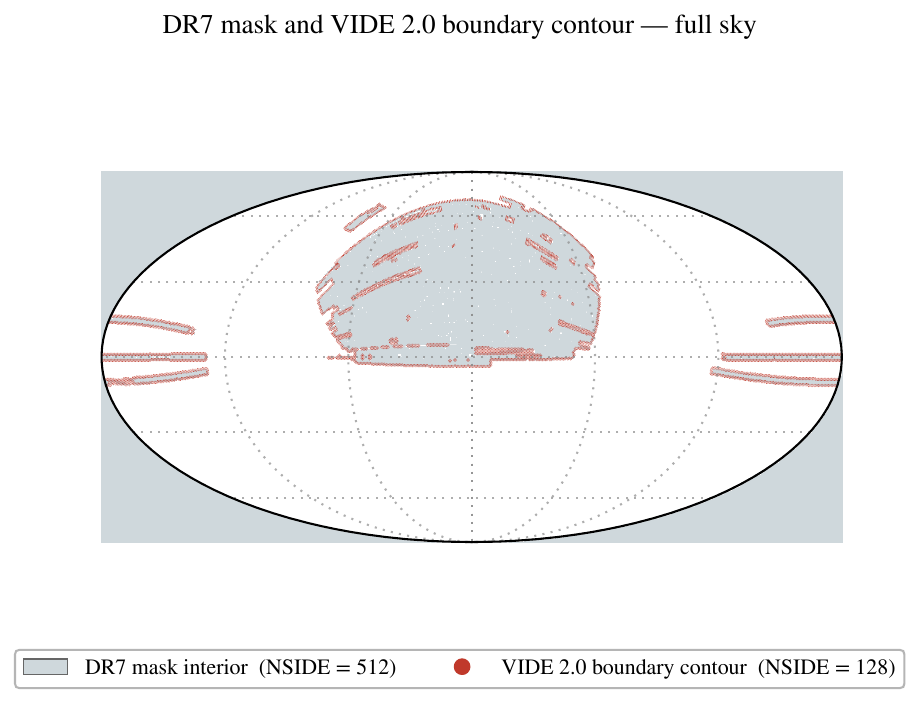}
    \plotone{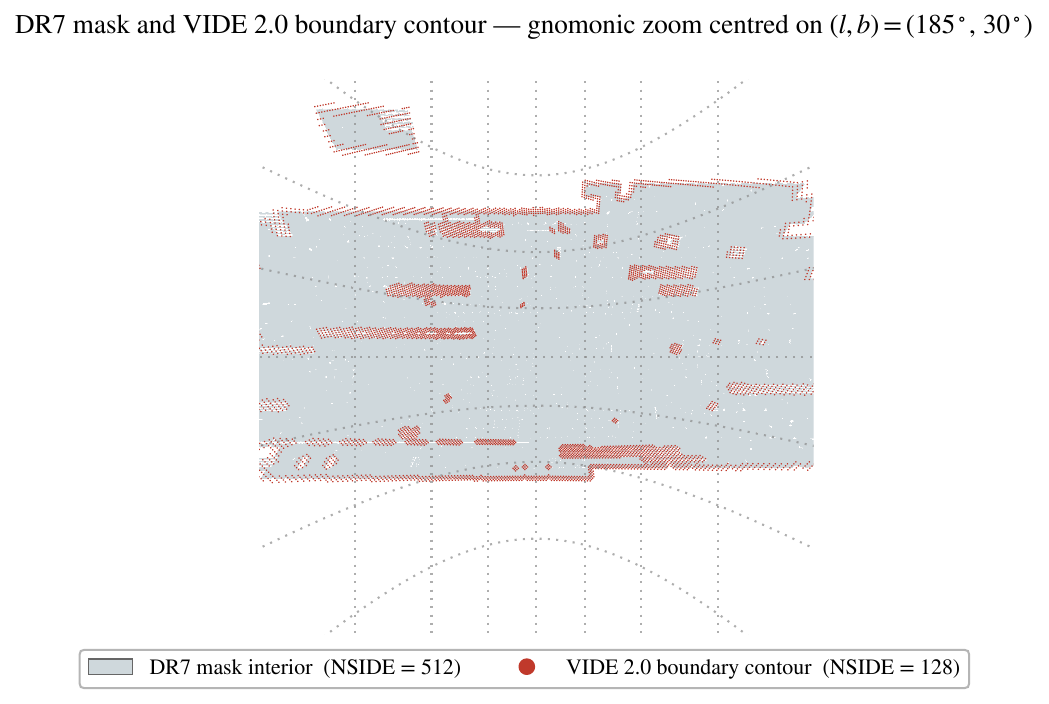}
  \caption{
           The first step in \vide's mask handling procedure, shown using
           a fiducial example drawn from the SDSS galaxy survey. Both
           panels overlay the input mask interior (gray) at its native
           resolution \nside~$=512$ on the boundary contour identified
           by \videtwo~(red) at \nside~$=128$, the latter chosen here
           for presentation clarity.
           \emph{Top:} full-sky Mollweide projection.
           \emph{Bottom:} gnomonic zoom centered on the densest region
           of the survey.
          }
\label{fig:mask}
\end{figure}

Since \vide~was originally developed in the context of the first void \ap~test, during
post-processing the default void catalog produced by \vide~implemented
additional filters, such as eliminating any void whose maximum extent (as
defined relative to its volume-weighted center) was greater than the distance
to the nearest boundary particle. This mitigated any mask-alignment biases.
Note that \vide~also produced a parallel catalog of \emph{all} voids, but this
catalog also included voids directly adjacent to mask boundaries. These cases were valid watershed basins representing underdense regions, but with derived information, such as extent, shape, and center, that could be affected by the mask they abutted or intersected.

Although this approach worked at the time and, indeed, delivered reliable \ap~
results and a bounty of other void information and follow-up studies, it suffers
from three drawbacks. 

The first is one of stochasticity. Since we place the boundary particles
randomly, different runs of \vide~on the same dataset can produce slightly
different void catalogs, as the underlying tessellation (i.e., topology)
changes for each realization of boundary particles. Because of this, small pinholes in the mask may not get a sufficient
density of boundary particles, leading \vide~to believe that these regions are
fairly sampled when in reality they are not.

The second is the introduction of another artificial parameter, the density (or equivalently nearest-neighbor distance) of
boundary particles, which violates the spirit of attempting to keep the void
finding approach flexible, universal, and as agnostic to the introduction of additional scales as possible.

Lastly, and most severely, large surveys covering substantial cosmological
volumes required an enormous number of boundary particles to adequately close
the survey domain. In many cases, these particles vastly outnumbered the
galaxies in the survey proper, leading to a significant loss of scalability,
performance, and ease of use.

For these reasons, for \videtwo~we targeted the boundary handling step as the
critical component for updating.

\subsection{Edge-galaxy flagging in \videtwo}
\label{sec:masks_v2}

Our new procedure, designed to alleviate these issues, is summarized in Algorithm~\ref{alg:edgeflag}. 

\begin{figure}
\renewcommand{\figurename}{Algorithm}
\setcounter{figure}{0}
\begin{algorithmic}[1]
\Require Galaxy positions $\{g_i\}$, survey mask $M$ at resolution $N_{\rm side}$, redshift bounds $[z_{\min}, z_{\max}]$
\Ensure Set of flagged edge galaxies $\mathcal{F}$

\State Run Voronoi tessellation on $\{g_i\}$ using \mbox{\textsc{voro++}}
\State Build KD-tree $\mathcal{T}$ from $\{g_i\}$
\State Identify boundary pixels $\{p_j\}$ from mask $M$
\State Compute minimum linear resolution $\delta$ from $N_{\rm side}$ at the survey's midpoint distance

\For{each boundary pixel $p_j$}
    \State Define ray $r_j$ along pixel center from $z_{\min}$ to $z_{\max}$
    \State Divide $r_j$ into redshift bins; refine angular sampling via \mbox{\textsc{HEALPix}} nesting to maintain spacing $\leq \delta$
    \For{each sample point $s$ along $r_j$ and any refined sub-rays}
        \State Query $\mathcal{T}$ for nearest galaxy $g_k$ to $s$
        \State $\mathcal{F} \leftarrow \mathcal{F} \cup \{g_k\}$
    \EndFor
\EndFor

\State Sample low-$z$ and high-$z$ caps at spacing $\delta$; flag nearest galaxies
\State Set Voronoi volumes of all $g \in \mathcal{F}$ to $\sim 0$
\State Proceed with watershed transform
\end{algorithmic}
\caption{\videtwo~Edge Galaxy Flagging}
\label{alg:edgeflag}
\renewcommand{\figurename}{Figure}
\end{figure}

As a first step, we replaced the underlying
tessellation engine, the \textsc{qhull} library~\citep{Barber2013}, with
\textsc{voro++}~\citep{Rycroft2009}. In addition to significant improvements in
performance, scalability, and portability (to be demonstrated below),
\textsc{voro++} provides a clearer relationship between the point set to be
tessellated and the bounding volume containing those points. In particular,
with \textsc{qhull} if a particle is near a boundary its tessellation is
ambiguous. In contrast, \textsc{voro++} explicitly includes the boundaries as
labeled faces of a given particle's Voronoi tessellation, if that boundary
intersects its local volume. While this does not remove the ambiguity of the volumes associated with galaxies near the edge of the survey region, it provides a zero-cost,
exact geometric determination of which galaxies have Voronoi cells
clipped by the domain boundary --- information that was fundamentally
unavailable in the previous scheme, where edge proximity could only be
inferred statistically from the random boundary particle distribution.

As for the angular survey bounds, in contrast to \videone's approach of
flooding mask edges with boundary particles, for \videtwo~we adopt an approach
based on identification of galaxies near the edges, flagging them, and
modifying the tessellation. This new approach maintains the spirit of treating
voids purely as morphological entities, and the impact of the mask as necessarily
morphological in nature as well.

This means that the user must clearly specify the survey mask in a way that
\vide~can understand. To do this, in addition to the raw mask file the user
must choose \nside, the desired \healpix~resolution for the internal
representation of the mask within \vide. We describe \videtwo's automatic
rule for selecting \nside~--- which eliminates it as a free parameter ---
in Section~\ref{sec:masks_nside} below. For the present discussion we
take it as given.

Once the mask properties are determined, the next step for \videtwo~is to
simply run the tessellation as-is. This ``raw'' tessellation contains mostly
useful density field calculations (for regions far away from any survey
boundaries) but also many pathological cases: Voronoi volumes that jump across
gaps in the survey, volumes that connect to the cubic bounding box containing
the survey volume, and so on.

To identify these pathological cases we drive rays down the pixel centers
defining the mask edges. We implement a memory-efficient linear search to find
all the nearest neighbor galaxies to any point on the line, down to a
resolution threshold determined by the comoving linear separation of
\healpix~pixels at the midpoint of the survey's comoving distance range.

\begin{figure}[htbp] 
    \plotone{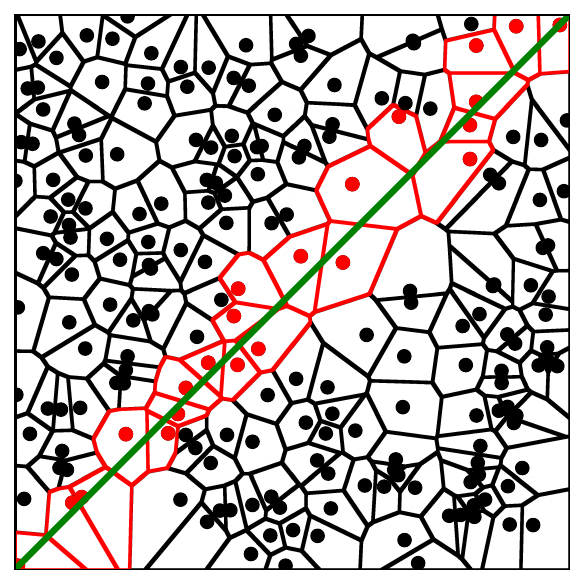}
  \caption{
      Cartoon depicting flagging technique. Shown is a Voronoi tessellation of a 2D grid of random points. The observer is at the lower left corner. A pinhole in the mask punches a radial ray through the survey, represented here as the green line. Galaxies ``nearest'' to this ray are flagged in red. We define ``nearest'' to mean any particle whose local Voronoi volume is intersected by the ray. The flagged galaxies have unreliable volumes and densities due to the presence of the pinhole.
         }
\label{fig:raysearchcartoon}
\end{figure}

Figure~\ref{fig:raysearchcartoon} shows a cartoon representation of this
nearest-neighbor searching. This approach of flagging along rays works for both
pinholes (as depicted) and survey edges.

At higher redshifts these rays can separate, allowing Voronoi volumes to
potentially ``spill'' outside the survey bounds. To account for this we divide
the survey volume into redshift bins and add higher-resolution rays using the
nested refinement capabilities of \healpix. We choose the nesting such that we
maintain a consistent linear separation between rays defined by \nside~at the
survey's midpoint distance throughout its entire volume.

Finally, we sample a discrete set of points at the low- and high-$z$ redshift
ends of the survey at the appropriate spacing so as to maintain the minimum
linear resolution determined by \nside.

The net result of this procedure is equivalent to discovering which galaxies
would have Voronoi volumes intersected by the outermost edges and internal holes
of the survey, and whose volumes would ``jump'' across disparate regions of the
survey. This makes their local density estimate unreliable and hence unsuitable
for the watershed transform. Flagging complete, we then take any edge galaxy
and set their Voronoi volumes to be (essentially) zero. When the watershed
transform proceeds to find basins and join them into voids, these edge-galaxies
act as infinitely-high walls that prevent the watershed from spilling outside
the survey bounds.

Importantly, \videtwo~distinguishes among three independent sources of
edge contamination, each recorded with a distinct flag value: (1)~galaxies
whose Voronoi cells are geometrically clipped by the tessellation domain
boundary, identified directly by \textsc{voro++}; (2)~galaxies nearest
the low- and high-redshift caps of the survey volume; and (3)~galaxies
nearest the angular mask edges, found via the ray-based search. This
per-galaxy provenance information is stored in the output catalogs,
enabling users to selectively filter based on the \emph{source} of
contamination --- for example, retaining voids near redshift boundaries
while excluding those affected by angular mask holes.

\subsection{Automatic Selection of NSIDE Resolution}
\label{sec:masks_nside}

\begin{figure*}[t]
  \plotone{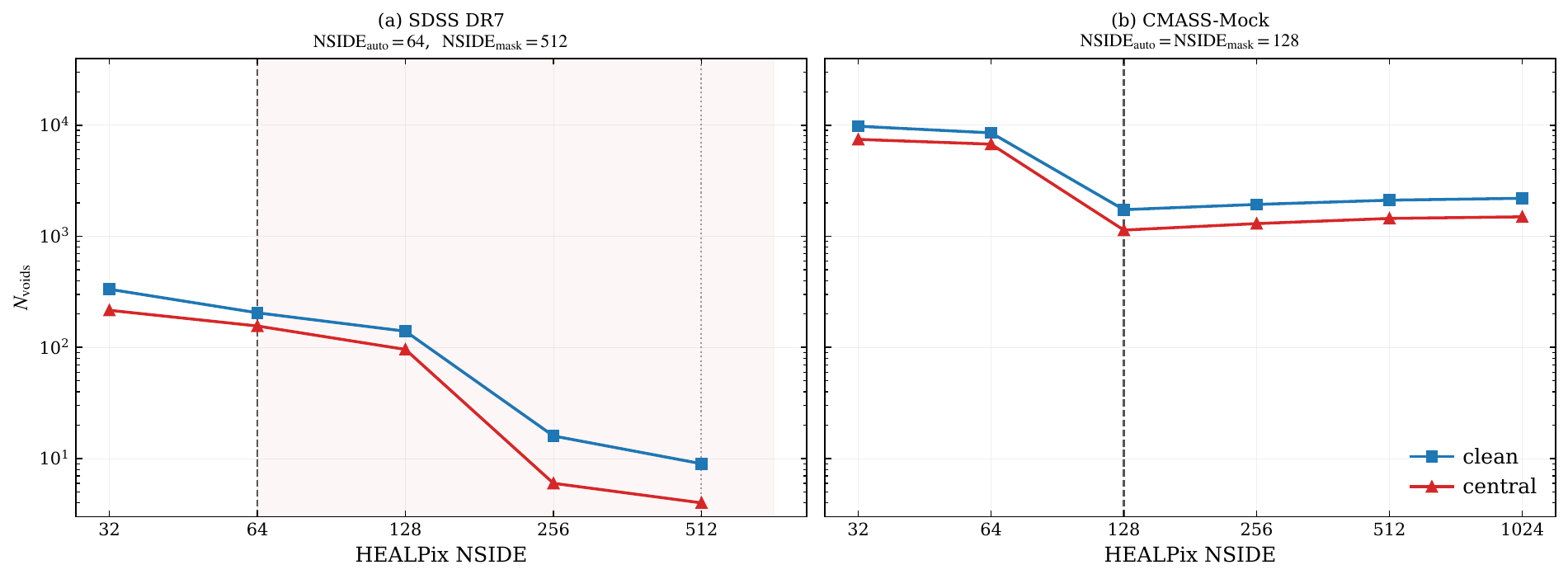}
  \caption{
    Number of voids identified by \videtwo~as a function of
    \healpix~\nside, separated into the two trustworthy catalog
    tiers: \emph{clean} voids (no edge-flagged member galaxies)
    and \emph{central} voids (no edge-flagged members and
    macrocenter farther from any survey edge than the void's
    maximum extent). \emph{Left:} SDSS DR7, whose mask has native
    resolution \nside~$=512$. \emph{Right:} a CMASS-like mock
    lightcone on the BOSS DR12 CMASS NGC footprint, whose mask has
    native resolution \nside~$=128$.
    Vertical dashed line: \nside~chosen by the auto rule of
    Equation~\ref{eq:nside_auto}; vertical dotted line:
    input mask's native \nside~when distinct from the auto
    value. The shaded red band on the left panel is the
    Mode~B regime where pinhole-piercing on the complex DR7
    mask collapses both catalog tiers by roughly two decades.
  }
  \label{fig:nside_auto}
\end{figure*}

Up to this point we have taken the \healpix~resolution \nside~at
which the mask is processed as given. In \videone~this was a free
parameter the user had to tune by hand, and in practice the choice
was non-obvious. Two failure modes constrain it. \emph{Mode~A} is
under-resolution: when the mask is coarser than the typical galaxy
separation, real survey edges are smoothed over, some edge-adjacent
galaxies escape flagging, and the \emph{clean} and
\emph{central} catalogs are inflated by spurious voids that should
have been flagged. \emph{Mode~B} is the opposite: when the mask is
much finer than the data and the mask carries non-trivial internal
topology (small interior pinholes, jagged outer edges, or both), each
finer subdivision of \nside~resolves smaller mask features that drive
volume-piercing rays through the survey interior, the watershed is
over-segmented, and the clean and central populations collapse
rapidly. Both modes are visible in Figure~\ref{fig:nside_auto}.

The empirical safe zone lies between these two failure modes,
bounded above by the mask's intrinsic resolution and below by the
data's density. We make this concrete in two steps. The mask side
is straightforward: read $\mathrm{NSIDE}_\mathrm{mask}$ directly from
the input \healpix~file. For the data side, we ask at what
\nside~the volume-weighted mean linear pixel resolution matches the
mean tracer separation. For a survey with comoving distance $D_C(z)$
between bounds $[z_\mathrm{min}, z_\mathrm{max}]$, this
volume-weighted mean is
\begin{equation}
  \langle r_\mathrm{pix} \rangle_V
    \;=\; \theta_\mathrm{pix}\;
          \frac{3}{4} \;
          \frac{D_C^4(z_\mathrm{max}) - D_C^4(z_\mathrm{min})}
               {D_C^3(z_\mathrm{max}) - D_C^3(z_\mathrm{min})},
  \label{eq:vw_resol}
\end{equation}
where $\theta_\mathrm{pix} = \sqrt{4\pi/N_\mathrm{pix}}$ is the
characteristic \healpix~pixel angle and
$N_\mathrm{pix} = 12\,\mathrm{NSIDE}^2$. Setting
$\langle r_\mathrm{pix}\rangle_V$ equal to the mean tracer separation
$\bar{s} = (V_\mathrm{survey} / N_\mathrm{tracers})^{1/3}$ and
rounding up to the nearest power of two yields the
data-density-matched value
\begin{equation}
  \begin{aligned}
    \mathrm{NSIDE}_\mathrm{data}
      &\;=\; 2^{\lceil \log_2(\sqrt{\pi/3}\,/\,\theta^\star) \rceil}, \\
    \theta^\star
      &\;=\; \frac{4\bar{s}}{3}\,
            \frac{D_C^3(z_\mathrm{max}) - D_C^3(z_\mathrm{min})}
                 {D_C^4(z_\mathrm{max}) - D_C^4(z_\mathrm{min})}.
  \end{aligned}
  \label{eq:nside_data}
\end{equation}
The auto rule is then simply
\begin{equation}
  \mathrm{NSIDE}_\mathrm{auto}
    \;=\; \min\!\bigl(
            \mathrm{NSIDE}_\mathrm{mask},\;
            \mathrm{NSIDE}_\mathrm{data}
          \bigr).
  \label{eq:nside_auto}
\end{equation}
Both terms are reported at run time so the user can confirm or
override the selection. We use the volume-weighted form in
Equation~\ref{eq:vw_resol}, rather than evaluating at the median
redshift, because a median-$z$ choice would under-represent the
larger-volume regions at high redshift where pixel sizes are
largest and the danger of under-resolution is greatest. The volume
weighting is the natural average for an observable --- the void
number function --- that integrates over the whole volume.

For the SDSS DR7 example used throughout the present paper, the
auto rule selects $\mathrm{NSIDE}_\mathrm{auto} = 64$
($\mathrm{NSIDE}_\mathrm{mask} = 512$,
$\mathrm{NSIDE}_\mathrm{data} = 64$); the data side wins because
the volume-weighted pixel resolution at \nside~$=64$ already matches
the mean tracer separation of $7.4\,$\hmpc. With this rule in place,
the only remaining input \videtwo~requires from the user is the
survey mask itself.

\subsection{Demonstration on SDSS DR7}
\label{sec:masks_demo}

Throughout this paper we exercise \vide~on three datasets: the SDSS DR7
sample packaged with \vide~as an exemplar, used in this section, plus
two Stage~III/IV-scale mock catalogs introduced for the demonstrations
of Section~\ref{sec:demos}. Their properties are summarized in
Table~\ref{tab:mocks}. We chose DR7 as the boundary-handling exemplar
because, while it is neither modern nor large by current standards, it
has a particularly complicated mask --- many internal holes and jagged
outer edges --- making it the ideal test case for the new procedures
developed above, as well as small enough to easily visualize for demonstration purposes.

\begin{table*}[t]
  \centering
  \begin{tabular}{lcc}
    \hline
    Property & AbacusSummit DESI Y1 ELG & MultiDark-Patchy DR12 NGC \\
    \hline
    Simulation
      & AbacusSummit
      & MultiDark Planck \\
    Galaxy sample
      & DESI Y1 ELG
      & BOSS LOWZ+CMASS LRG \\
    Footprint
      & DESI Y1 NGC+SGC 
      & BOSS DR12 NGC \\
    Redshift range
      & $0.8 \le z \le 1.6$
      & $0.2 \le z \le 0.75$ \\
    Sky fraction
      & $0.22$ ($\sim 9{,}100\,$deg$^{2}$)
      & $0.18$ ($\sim 7{,}400\,$deg$^{2}$) \\
    Galaxies per realization
      & $1.21 \times 10^{7}$
      & $9.2 \times 10^{5}$ \\
    Mean tracer separation
      & $\sim 12\,h^{-1}\,$Mpc
      & $\sim 17\,h^{-1}\,$Mpc \\
    Realizations used
      & $25$
      & $200$ \\
    \hline
  \end{tabular}
  \caption{Mock suites used for the demonstrations of
    Section~\ref{sec:demos}. In the text we refer to them as the
    Abacus ELG and Patchy mocks.}
  \label{tab:mocks}
\end{table*}

\begin{figure*}[tp] 
    \plottwo{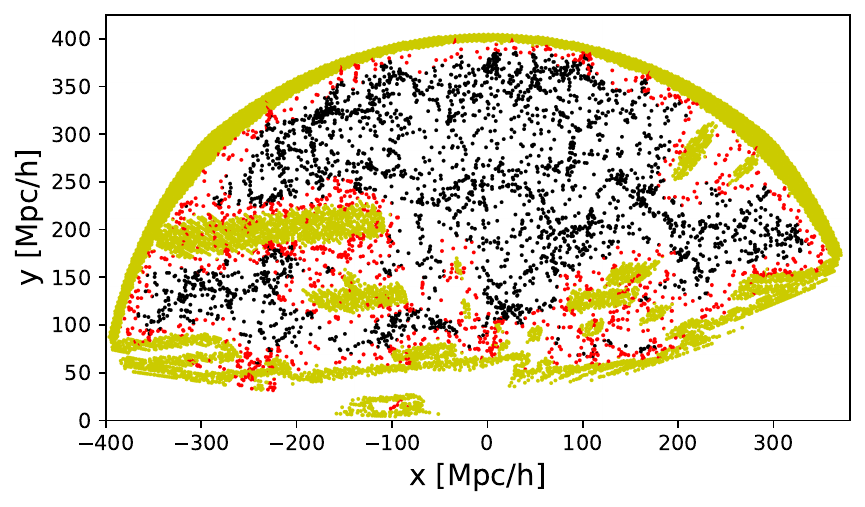}{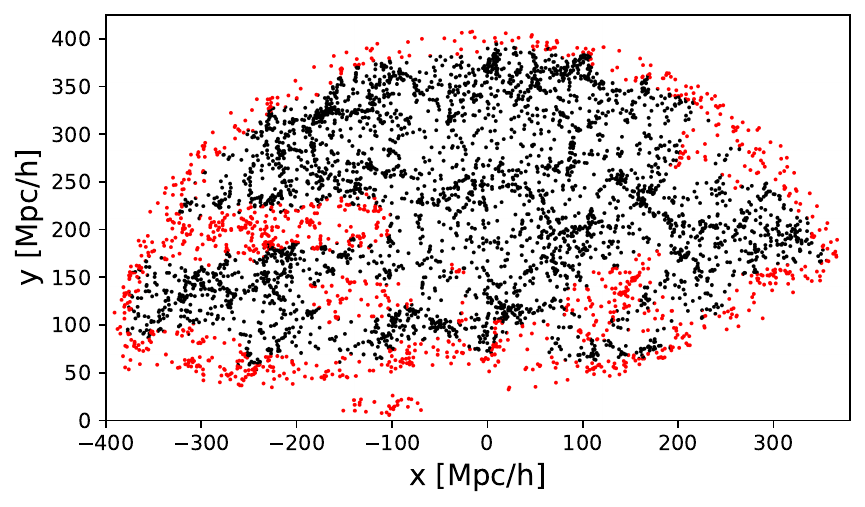}
  \caption{
           Comparison between old and new survey boundary handling procedures. Shown is a slice taken from an example survey (in particular, the SDSS DR7 dataset packaged as an exemplar with \vide). We deliberately have taken a slice through the survey to showcase both boundary edges and internal holes. Galaxy positions are converted to \hmpc~assuming a fiducial \lcdm~cosmology. 
         \emph{Left}: \videone~scheme, with galaxies nearest the slicing plane shown as black points, inserted boundary particles as yellow points, and galaxies adjacent to boundary particles (and hence removed from the watershed step) as red points.
       \emph{Right}: \videtwo~scheme, with galaxies nearest the slicing plane shown as black points, and flagged galaxies nearest to boundaries as red points.
         }
\label{fig:boundpart}
\end{figure*}

Figure~\ref{fig:boundpart} highlights the differences between the \videone~and
\videtwo~approaches to boundary handling in a slice through the DR7
volume-limited subsample at $z = 0$--$0.15$.

In this example case, of the 164,647 original galaxies in the sample,
\vide~flagged 42,017, or roughly 25\% of them, as edge galaxies. This large
fraction is due to the relative sparsity of the surveyed galaxies and the
aforementioned complexity of the mask. This is roughly equivalent to the number of edge galaxies flagged in \videone.

\section{Performance \& Scaling}
\label{sec:scaling}

The benefits to this new approach are manifest. For one, it is extremely
lightweight. With no additional particles added, there is no additional cost to
running \vide~with larger surveys, aside from the expected cost of dealing with
greater numbers of galaxies.

Secondly, this method is fully deterministic. \videone~had two
sources of stochasticity: the random placement of boundary particles
to flood the survey edges, and the variable sampling of pinholes
when the mask resolution was insufficient. Two consecutive
\videone~runs on the same input therefore produced slightly
different catalogs, with run-to-run scatter in core densities,
density contrasts, and membership for voids near the survey edges.
\videtwo~places no random particles, and given identical mask,
\nside, and tracer positions it identifies the same edge galaxies
(Algorithm~\ref{alg:edgeflag}) before passing them to the (deterministic)
\textsc{voro++} tessellator. The pipeline is therefore reproducible
at the bit level: re-running on the same input yields SHA-256
identical catalog files, and the single-domain and KD-tree-split
configurations agree bit-for-bit at the recommended buffer width.

Combined with the auto-\nside~rule of
Section~\ref{sec:masks_nside}, the boundary-handling step requires
no further tuning: the user supplies only a tracer catalog and a
mask.

Lastly, the new approach is much faster and exhibits better scaling behavior
than the mask handling scheme in \videone. The additional computational cost
comes from the need to, essentially, cross-correlate every boundary pixel with
every galaxy. However, we achieve significant speedups by placing the galaxies
on a KD-tree, which makes the nearest-neighbor search at each
reference point along a ray an $\bigO(\log n)$ operation. We tested several schemes for searching along the ray,
including a bisection search, but found that a simple linear stepping provided
the most robust, memory- and compute-efficient approach. The bisection search remains available via a user-selectable option should it prove useful in other scenarios.

While the search for edge-adjacent galaxies does incur an additional
computational cost, it relieves \vide~from having to consider a substantial
number of boundary particles in the tessellation, which dominate the computational cost, and the watershed transform. In some
cases, this made \videone~outright unusable, with so many particles required
that even memory-generous systems were overwhelmed.

Additionally, \textsc{voro++} is much more lightweight than \textsc{qhull}.
Even though the latter is more widely used, it is much more general, meant for
tessellations in any number of dimensions. In our experience, \textsc{voro++}
has proven to be roughly 30--50\% faster and require significantly less memory
in the Voronoi tessellation step.

To test the scaling behavior we ran the full pipeline on mock0, one
realization of the AbacusSummit DESI~Y1 \texttt{ELG\_LOP\_complete}
mocks (the Abacus ELG mocks of Table~\ref{tab:mocks}). Mock0
contains $1.21 \times 10^{7}$ galaxies, and we also ran seven
uniformly random subsamples of it down to $1.25 \times 10^{5}$
tracers. We left
every parameter at its default so that the curves reflect what a
user gets without any tuning: a KD-tree subdomain split with two
domains, automatic thread allocation, and the auto-\nside~rule,
which selected \nside$=64$, $128$, and $256$ for the small,
intermediate, and large samples respectively. All runs used
eight processes on an Apple M2 MacBook Air with 8\,GB of memory.

\begin{figure*}[tp]
    \plottwo{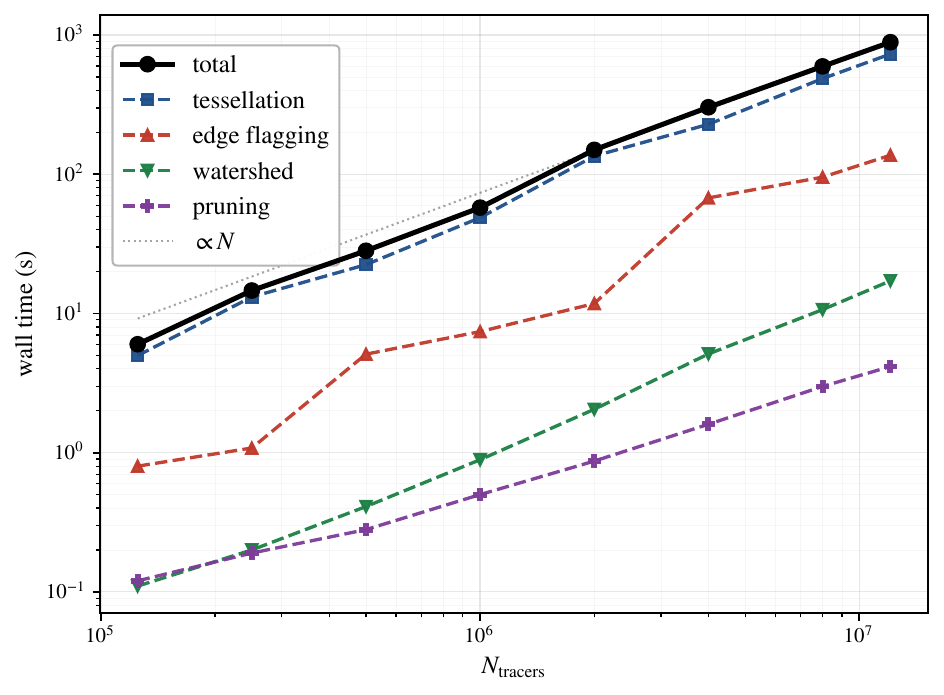}{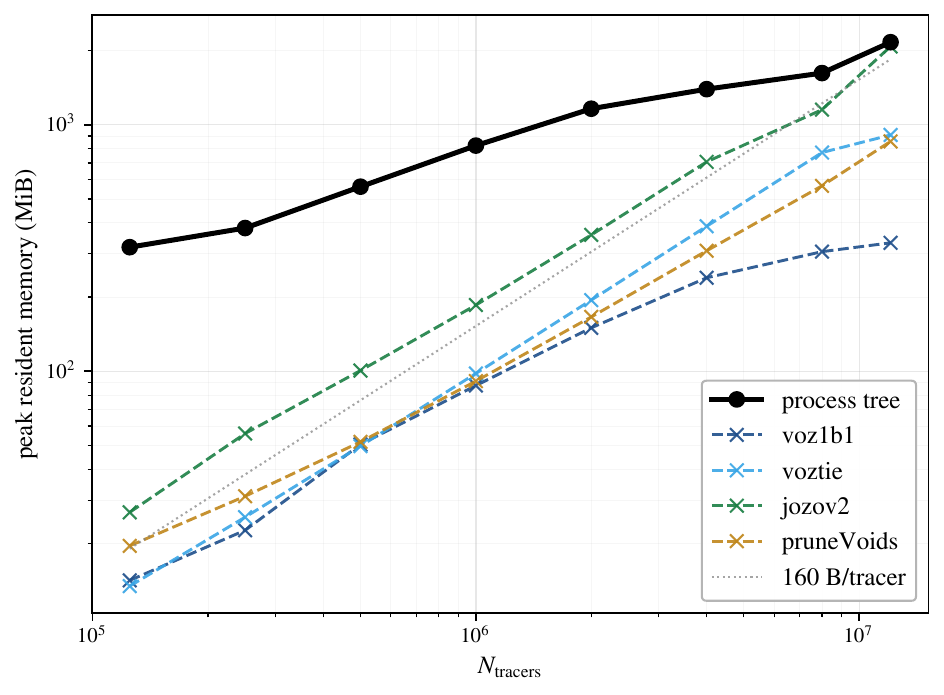}
  \caption{Scaling performance of \videtwo~on the Abacus ELG
           mock0 catalog and uniform random subsamples of it, on an Apple M2 MacBook Air
           (8\,GB). \emph{Left:} wall time per pipeline stage versus
           tracer count. The dotted line marks $\bigO(n)$ scaling.
           \emph{Right:} peak resident memory of the full process
           tree and of each pipeline binary; the dotted line is the
           asymptotic $160\,$B per tracer set by the watershed
           step.}
\label{fig:scaling}
\end{figure*}

Figure~\ref{fig:scaling} shows the result. Wall time and peak
memory both scale near-linearly in $N$ across two orders of
magnitude in catalog size. The Voronoi tessellation dominates the
wall clock budget at every size; edge flagging is the next-largest
contributor, with visible bumps each time the auto-\nside~rule
increases the mask resolution. The watershed and pruning steps
together account for only about 2\% of the wall time.
\textsc{jozov2} tracks an asymptotic memory footprint of
$\sim 160\,$B per tracer, with the full process tree at
$\sim 250\,$B per tracer. Tessellation aside, all components
approach $\bigO(n)$ scaling at large $N$ once the KD-tree of
galaxy positions is in place.

For comparison: \videone~required roughly 10 minutes to process
the SDSS DR7 sample of $\sim 1.6 \times 10^{5}$ galaxies, while
\videtwo~runs the same problem in a handful of seconds. At the other
end, the full $1.21 \times 10^{7}$-galaxy Abacus ELG
mock --- a catalog two orders of magnitude larger than DR7 ---
completes in $\sim 15$ minutes and peaks at $\sim 2.2$\,GiB on the
same laptop. \videone~could not run a catalog of even $10^{6}$
galaxies on this hardware: it crashed from memory exhaustion after
$\sim 30$ minutes, well before the watershed step. CMASS-class
samples ($\sim 10^{6}$ galaxies) and DESI~Y1 ELG-class samples
($\sim 10^{7}$) are therefore inaccessible to \videone~on a laptop
and tractable with \videtwo~in minutes.

We also made several other changes as part of the upgrade.

\paragraph{Per-tracer weights.}
\videtwo~accepts an optional weights file as a sample-level input.
After the Voronoi tessellation returns a per-tracer cell volume
$V_{i}$, the watershed input volume is set to
$V_{i}^{\mathrm{ws}} = w_{i}^{-1} V_{i}$. The mechanism is decoupled
from edge flagging --- edge-flagged tracers override the weight by
forcing $V_{i}^{\mathrm{ws}}$ to a near-zero wall value, so weights
cannot interact with boundary handling --- and an all-ones weights
file produces a bit-identical catalog to an unweighted run.

\paragraph{Auto-resolved defaults.}
In addition to the auto-\nside~rule of
Section~\ref{sec:masks_nside}, \videtwo~auto-resolves the subdomain
decomposition mode (a KD-tree split for observational data, a
regular cubic split for simulations) and the per-subdomain buffer
width ($0.30$ for the KD-tree mode, $0.10$ for the regular grid). With
two domains, the KD-tree split delivers a $\sim 3\times$ speedup on the
tessellation while reproducing the single-domain catalog
bit-for-bit. Each auto-resolved choice is written to the run log
and can be overridden explicitly.

\paragraph{Build modernization.}
The original \vide~build relied on a hand-rolled \texttt{setup.py}
that drove CMake through a custom build extension and rebuilt every
internal C/C++ dependency at install time. On Python $\geq 3.12$
that path is broken, since \texttt{distutils} has been removed from
the standard library. \videtwo~is built with
\textsf{scikit-build-core} via \texttt{pyproject.toml}; we removed
\textsc{Boost}, \textsc{NetCDF}, \textsc{Eigen}, and \textsc{libSDF}
as build-time dependencies, copied the seven \textsc{CosmoTool}
headers that \vide~actually uses directly into its source tree, and
defaulted the remaining external libraries
(\textsc{HEALPix}, \textsc{cfitsio}, \textsc{GSL}, \textsc{voro++},
\textsc{HDF5}) to system packages. A clean \texttt{pip install vide}
now takes roughly two minutes on an Apple Silicon laptop with
Homebrew.

\paragraph{Other improvements.}
Internal pipeline I/O has moved from \textsc{NetCDF} to
\textsc{HDF5}, giving roughly an order-of-magnitude speedup on the
data-preparation stage. The per-tracer edge-flag field now carries
a small integer recording which of three independent
boundary-detection mechanisms (\textsc{voro++}'s native wall
detection, the redshift-cap ray-march, the angular-mask ray-march)
flagged each tracer, allowing downstream users to retain or exclude
voids by edge type. A lightweight process-tree poller records
per-stage peak resident memory in the run log alongside wall-clock
timing, and a \texttt{pytest}-based regression harness compares
end-to-end pipeline outputs against locked hash manifests on every
commit.

\section{Impacts on Void Catalogs}
\label{sec:impacts}

We first developed \vide~to apply watershed
techniques to real galaxy surveys and to identify voids in a
consistent, coherent fashion at scale in observations. For this reason, we
took an exceptionally cautious approach to boundary handling, wishing only to
include voids that were ``far'' from any survey edges, redshift boundaries, and
internal holes.

The choices proved prudent, but now with over a decade of experience with voids
in surveys, we can afford to take a more precise approach --- we can
get closer to the edge and still understand how the watershed algorithm is
behaving and the kinds of voids it produces. 

\begin{figure}[htbp] 
    \plotone{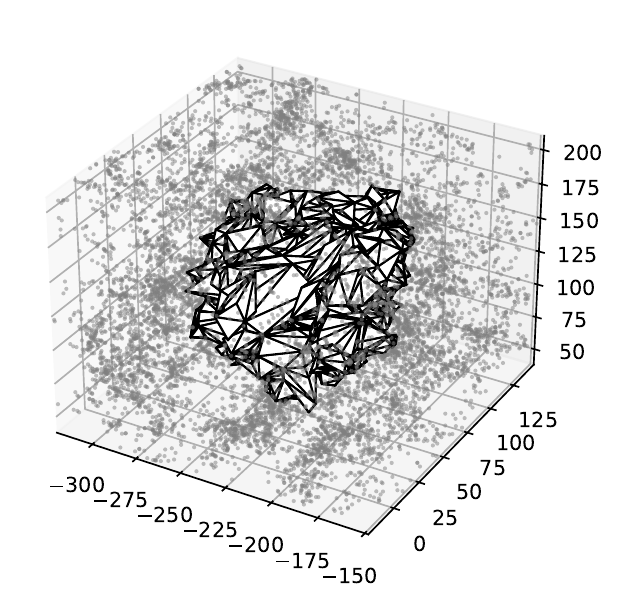}
  \caption{
         A typical large void identified with~\videtwo, with the cosmic web providing the bounding volume for the complex shape of the interior void basin.
         }
\label{fig:voidwireframe}
\end{figure}

\begin{figure}[htbp] 
    \plotone{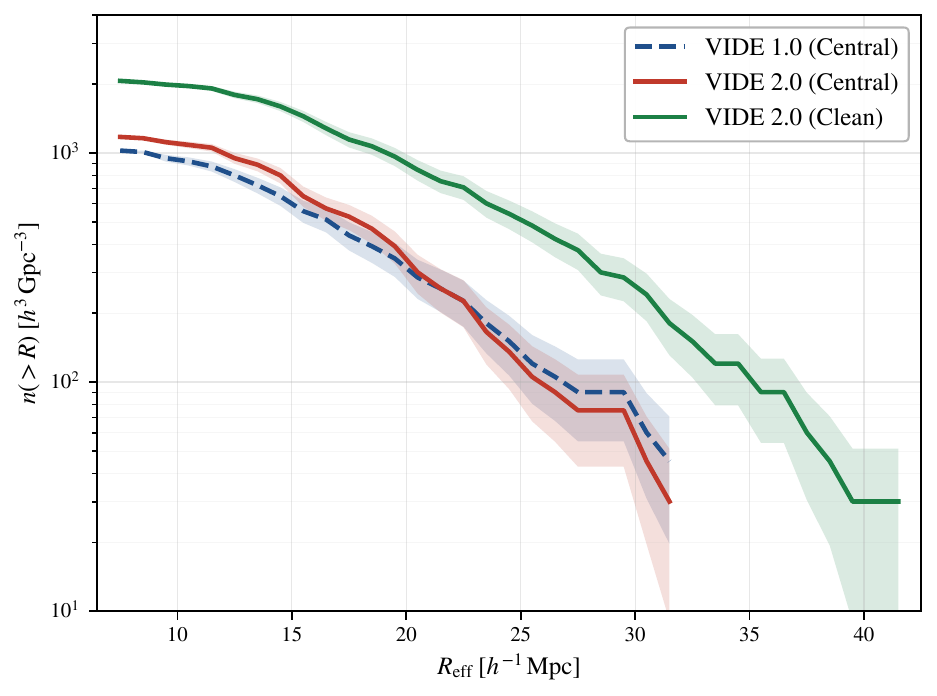}
  \caption{
      Comparison of void number functions between \videone~and \videtwo~catalogs in the SDSS DR7 galaxy sample. Catalogs are the \emph{central} voids, which are determined to be as far away from the boundaries as possible and minimally biased for \ap\ analysis, along with the full \emph{clean} set from \videtwo. The new boundary handling procedure eliminates the appearance of small, spurious voids near the boundaries, and allows for the persistence of large voids.
          }
\label{fig:numberfunc}
\end{figure}

The
combination of Voronoi tessellation and watershed algorithm explores the
complex, irregular bounding surface of the voids, up to the ridge-lines defined
by the spine of the cosmic web.
Figure~\ref{fig:voidwireframe} shows a typical large void identified with
\videtwo. Voids this large were usually discarded by the conservative
\emph{central} cut of \videone: in our SDSS~DR7 sample the \videtwo~\emph{clean}
catalog contains four times as many voids with $R_{\rm eff} > 25$~\hmpc~as the
\videone~\emph{central} catalog (32 versus 8; Figure~\ref{fig:numberfunc}).

\begin{figure*}[tp] 
    \plottwo{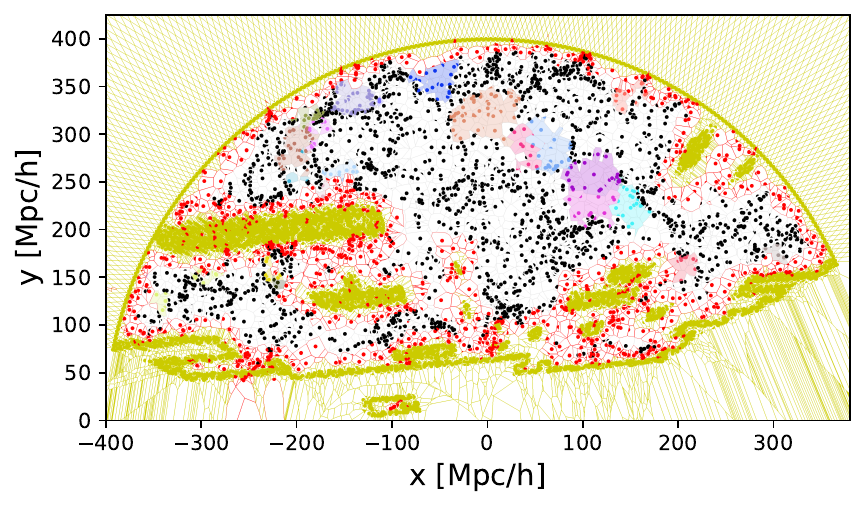}{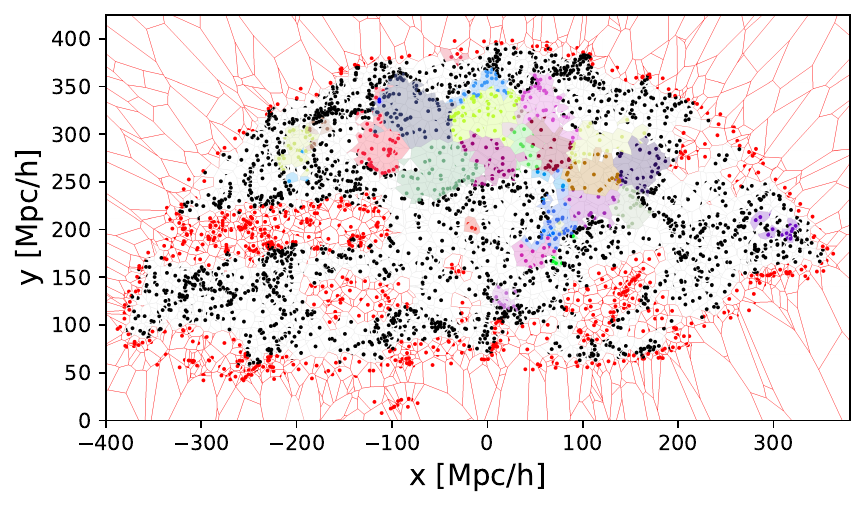}
  \caption{
      Comparison between \videone~\emph{central} voids (left) with \videtwo~\emph{clean} voids (right) for our fiducial test case. Inserted boundary particles and the edges of their Voronoi cells are shown in yellow, edge galaxies and their Voronoi cells in red, and regular galaxies in black. Note that this is a map of galaxies nearest the slicing plane, but for clarity we only display galaxies if they happen to lie within their corresponding Voronoi cell on this plane. In other words, this is not the Voronoi diagram of a 2D distribution of points, but a slice through the full 3D distribution.
      We overlay on this slice the identified voids, where we shade the Voronoi cells and member galaxies with random colors.
      The new boundary handling approach preserves voids in the central region of the survey, while preventing the formation of spurious small voids while also permitting the survival of large ones.
  }
\label{fig:slice_overview}
\end{figure*}

Indeed, we find significant overlap between void populations found with
\videone~and \videtwo, especially in the central regions of the survey, far
from any edges and holes, as shown in Figure~\ref{fig:slice_overview}. In this
figure, we compare the \emph{central} catalog of voids in \videone~(which
excludes any voids whose maximum extent from the macrocenter is farther than
the nearest boundary particle) to the \emph{clean} catalog of voids in
\videtwo, which excludes any void that explicitly contains a flagged edge galaxy as
a member. Unlike the geometric heuristic employed in~\videone, which
compared a void's maximum extent to the distance to the nearest
stochastically-placed boundary particle, this classification is exact: a
void is contaminated if and only if it contains an edge galaxy, with no
dependence on random realizations. This precision allows~\videtwo~to
retain significantly more voids near survey boundaries while maintaining
rigorous contamination control.

As efficient as the new mask handling procedure is, it does change the overall
topology of a given survey volume. In a sense, \vide's understanding of the
nature of the boundary is now different, which impacts the final population of
voids. Figure~\ref{fig:numberfunc} highlights these differences. 

\begin{figure*}[tp] 
    \plottwo{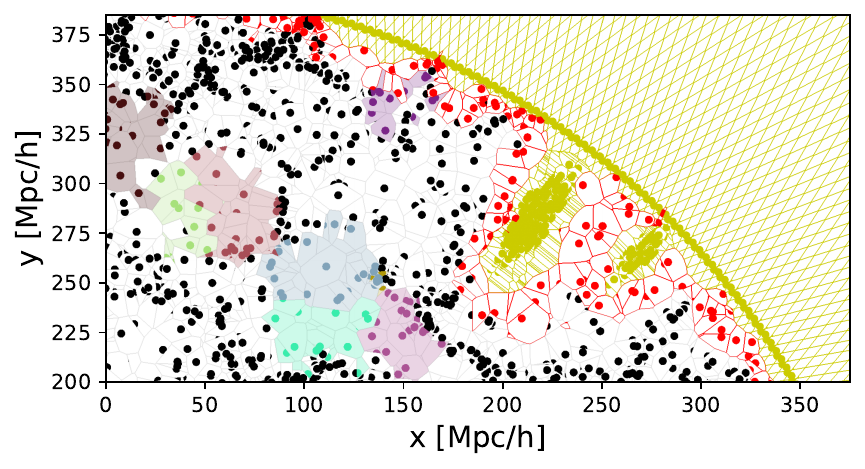}{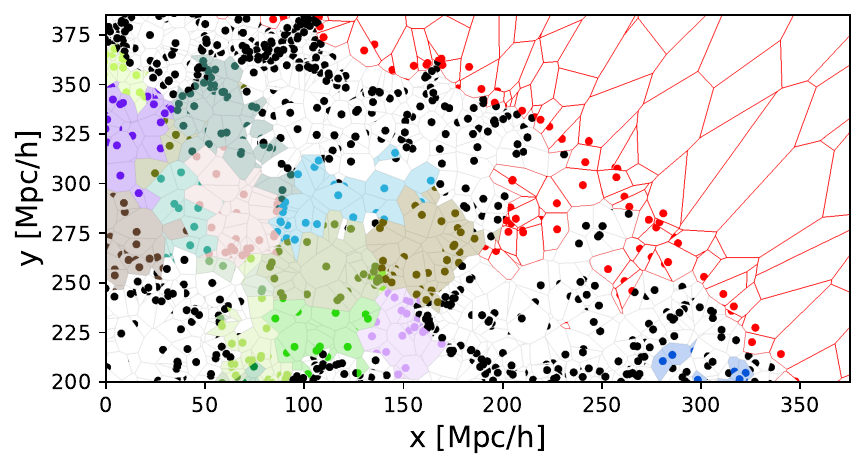}
    \caption{Same as Figure~\ref{fig:slice_overview} --- \videone~\emph{central} voids (left) compared to \videtwo~\emph{clean} voids (right) --- zoomed into the upper-right quadrant of the slice. Note the presence of a spurious small void near the edge at approximate coordinates $(x=150, y=325)$~\hmpc~in \videone~due to boundary effects, which are  handled more gracefully by \videtwo. In addition, \videtwo~allows many more voids to survive closer to the edge due to less aggressive filtering schemes.
    }
\label{fig:voidremoval}
\end{figure*}

While overall~\videtwo~identifies fewer small voids, this is a feature,
not a limitation. In~\videone, the randomly-placed boundary particles
introduced spurious density fluctuations that the watershed transform
interpreted as real voids, creating fictitious basins that do not
correspond to genuine underdensities in the galaxy distribution. These
artifacts preferentially appear at small scales, roughly on the order of
the mean galaxy separation, where the boundary particle density is
comparable to the tracer density. \videtwo~eliminates this entire class
of spurious detections, since no artificial particles
enter the tessellation. We highlight this effect in
Figure~\ref{fig:voidremoval}, which contrasts the much cleaner boundary
management in~\videtwo.

In contrast, the aggressive cleaning approach in \videone~removes many large
voids, as they might introduce a bias since they are near edges and holes. With
\videtwo, we find excellent agreement at large void radii when we impose a
similar severe restriction to mimic the \emph{central} catalog behavior.
Because the v2 reference point is an edge-flagged galaxy --- inside the
survey by definition --- rather than a mock particle placed outside, the
cut is slightly more restrictive, and most aggressive at the largest void
radii. Relaxing to the \emph{clean} catalog gives a much larger population
across almost all sizes.

\begin{figure}[htbp] 
    \includegraphics[width=\columnwidth]{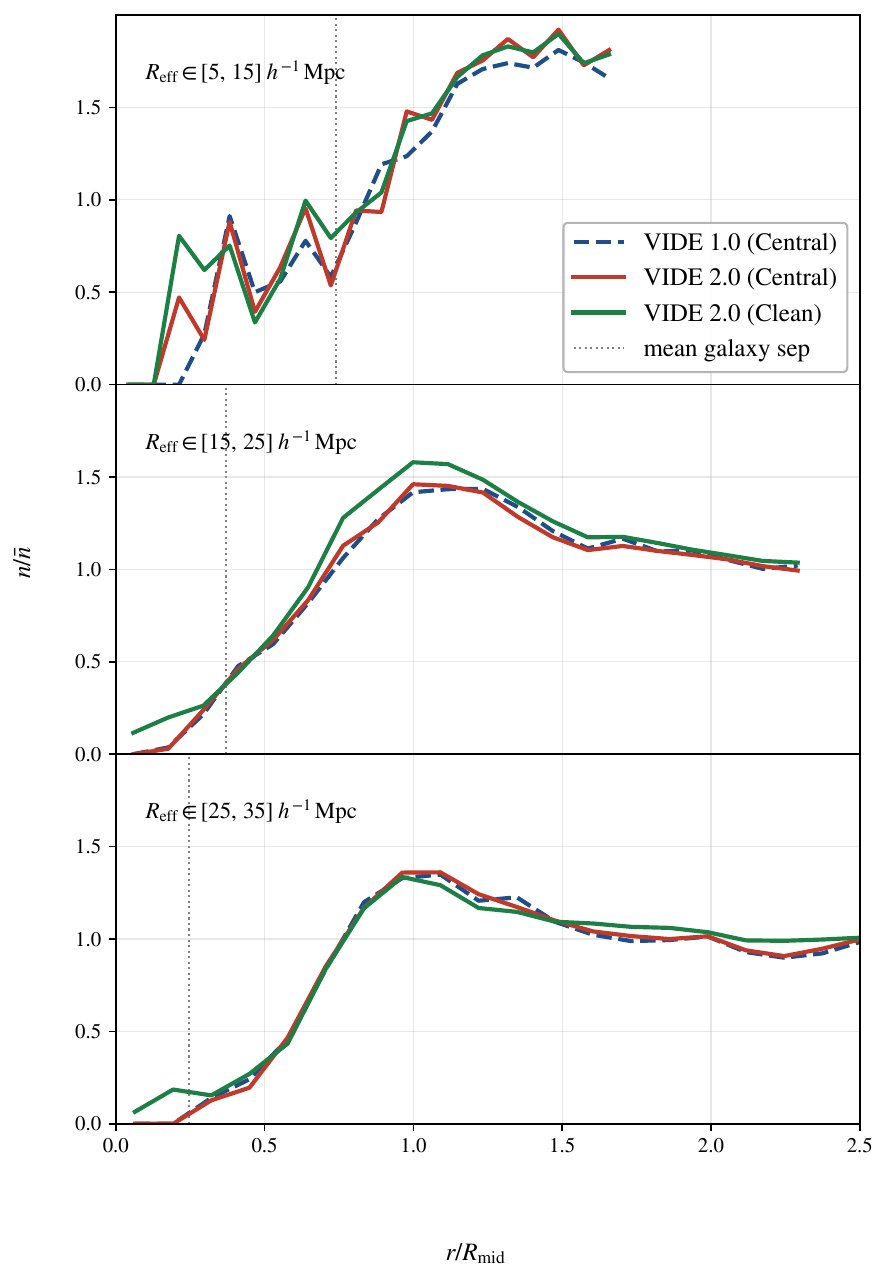}
  \caption{Comparison of radial density profiles for stacks of various void sizes. Distances are scaled by the midpoint radius $R_{\rm mid}$ of each void-size bin. The vertical dotted line indicates the mean galaxy separation in the sample, $\sim 7.4$~\hmpc. Note the excellent agreement for larger voids; the smallest voids, with radii near the mean galaxy separation, give noisy profiles in both versions.}
\label{fig:rprofiles}
\end{figure}

That said, \videtwo~still preserves critical void statistics. For example, the
radial density profile, which is equivalent to the void-galaxy
cross-correlation~\citep{Hamaus2016}, of stacked voids within given size bins
(Figure~\ref{fig:rprofiles}) remains nearly identical for voids larger than
$15$~\hmpc. This is perhaps
unsurprising. For these void sizes, most of the voids are away from the
boundaries and are exactly the same in the two catalogs. They are still
identified underdense regions in the galaxy distribution. For the smallest
voids ($5$--$15$~\hmpc), whose radii approach the mean galaxy separation, the
profiles from both versions are noisy and agree only within that noise.

\begin{figure}[htbp] 
    \plotone{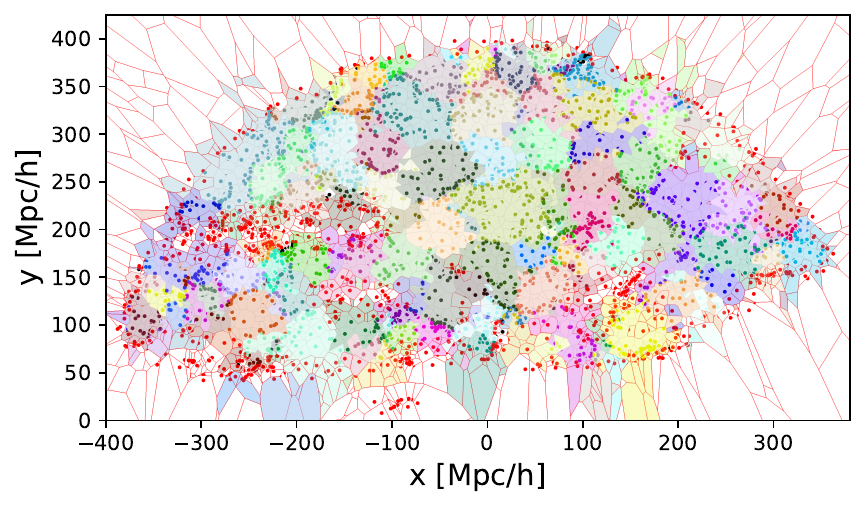}
  \caption{All voids, including edge-contaminated ones, identified with \videtwo. Note that Voronoi cells near the boundaries are ambiguously determined, but the regions interior to them contain valid underdensities.}
\label{fig:slice_allvoids}
\end{figure}

For completeness we show the \emph{all} catalog of voids identified in
\videtwo~in Figure~\ref{fig:slice_allvoids}. As expected, the universe is mostly
void~\citep{Douglass2023}, with watershed basins filling up the entire survey
volume. Despite the artificial high densities for any flagged galaxies, many
underdense regions approach the survey boundaries, where they are 
truncated by the lack of data.

Since voids have many uses across many analyses, \videtwo~classifies
every basin at pruning time. A void is \emph{contaminated} if any
of its member galaxies is edge-flagged, and \emph{central} if it
is clean (i.e., not contaminated) \emph{and} its maximum extent
from the macrocenter is shorter than the distance to the nearest
edge-flagged galaxy. The classification is deterministic, depending
only on the per-tracer edge flags and the macrocenter geometry. The
pipeline produces four standard catalogs:

\begin{itemize}
    \item \emph{all}: every watershed basin. This catalog is
      appropriate when only the location of an underdensity
      matters, as in weak-lensing stacks or cross-correlations with
      diffuse high-energy backgrounds.
    \item \emph{clean}: voids with no edge-flagged member galaxies.
      Because these voids are bounded entirely by unflagged tracers,
      their core densities, density contrasts, and adjacency-derived
      properties carry no imprint of the survey geometry. Use it for
      void-galaxy cross-correlations, ellipticity distributions, and
      alignments with large-scale structure.
    \item \emph{central}: a strict subset of \emph{clean} in which
      the void's maximum extent fits inside the survey on every
      side. Use it for radial density profiles, void lensing,
      \ap~tests, and other measurements keyed to the macrocenter.
    \item \emph{tree}: the hierarchical merger of adjacent basins
      into super-voids. This is not a primary science catalog ---
      it duplicates \emph{all} --- but it is useful for constructing
      void merger histories and for identifying broad, shallow
      super-voids.
\end{itemize}

On SDSS DR7 the \emph{clean} and \emph{central} views contain
$137$ and $78$ voids; on the $1.21 \times 10^{7}$-galaxy
Abacus ELG mock0 the corresponding counts are
$4.2 \times 10^{4}$ and $3.5 \times 10^{4}$.

In addition to these catalog-level classifications,~\videtwo~stores
per-galaxy edge flags indicating whether each tracer was flagged and, if
so, the source of contamination (domain boundary, redshift cap, or
angular mask edge). This granularity is new to~\videtwo~and enables
analyses that were previously impossible; for example, retaining
edge-adjacent voids for lensing cross-correlations while excluding
specific flagged galaxies from local density estimates, or studying how
different sources of survey incompleteness impact void properties
independently.

\section{Demonstration of Stage IV Readiness}
\label{sec:demos}

In this section we run \videtwo~on the two mock suites of
Table~\ref{tab:mocks}, generating the kinds of derived statistics a
Stage~IV analysis would need. We do not attempt a cosmological
measurement here; our goal is to show that the catalogs are ready
for one.

We ran \videtwo~over the first 200 realizations of the Patchy
mocks~\citep{Kitaura2016_PatchyDR12, RodriguezTorres2016_BOSS_HOD},
each containing $\sim 9.2 \times 10^{5}$ galaxies at $z \in [0.2,
0.75]$. 

We did the same for all 25 realizations of the Abacus ELG
mocks~\citep{Maksimova2021_AbacusSummit, Hadzhiyska2022_AbacusHOD,
DESI2024_Y1mocks}, each carrying $1.21 \times 10^{7}$ tracers at
$z \in [0.8, 1.6]$. Both ensembles used the auto-resolved defaults
of Section~\ref{sec:masks_nside} and the catalog-tier flags of
Section~\ref{sec:impacts}, and required no parameter retuning between
the BOSS-class and DESI-class samples.

The lack of stochasticity has direct consequences for ensemble work.
In \videone, the random boundary particles gave every catalog some
run-to-run scatter (Section~\ref{sec:scaling}). Across a mock
ensemble, that scatter adds a finder-specific term to the measured
covariance, on top of the cosmic variance we actually want to
measure. Because the term is independent between realizations, it
does not stall the convergence of the covariance estimate; instead
it inflates the converged value, and no statistic built from one run
per realization can separate it from cosmic variance. \videtwo~removes
the term entirely: re-running any realization reproduces its
catalog bit for bit, so the scatter across realizations comes
entirely from the realizations themselves.

\begin{figure}[htbp]
    \plotone{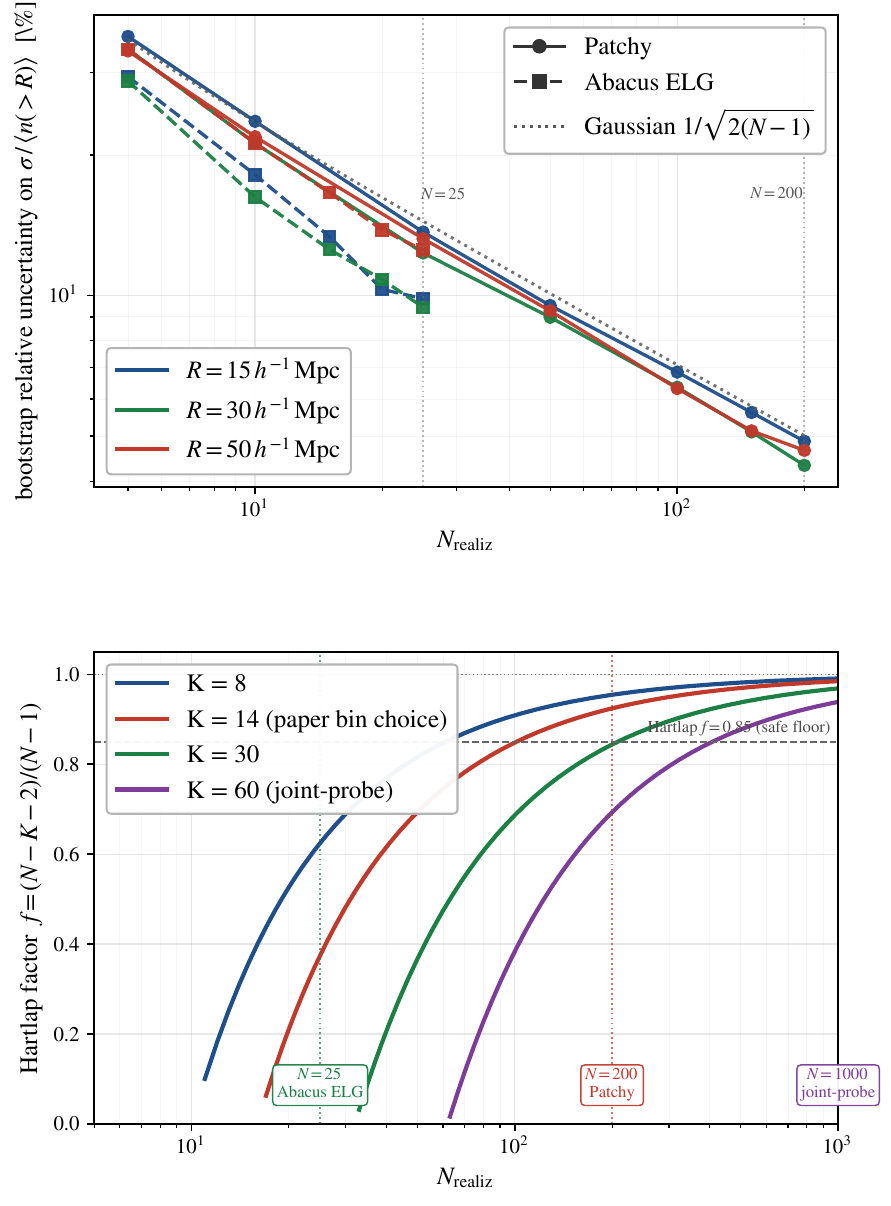}
  \caption{Convergence diagnostics for the cosmic-variance
           covariance. \emph{Top:} bootstrap relative uncertainty on
           the diagonal estimator $\sigma\,/\,\langle n(>R)\rangle$,
           resampling $N$ realizations with replacement, as a
           function of $N$, at three
           representative radii ($R = 15, 30, 50\,h^{-1}\,$Mpc).
           Solid lines with circle markers: Patchy (200
           realizations); dashed lines with square markers: Abacus
           ELG (25 realizations). The estimator's central value is
           essentially flat in $N$ and is omitted. The dotted line is
           the Gaussian expectation $1/\sqrt{2(N-1)}$ for a sample
           standard deviation, and the last point of each curve is
           the uncertainty for the full ensemble. \emph{Bottom:} Hartlap precision-matrix bias
           factor $f = (N - K - 2)/(N - 1)$
           \citep{Hartlap2007} as a function of $N$
           for several data-vector lengths $K$, with operating-point
           annotations at the ensemble sizes used here ($N = 25$ for
           Abacus ELG, $N = 200$ for Patchy) and a joint-probe
           extrapolation ($N = 1000$). The horizontal dashed line is
           the conventional $f = 0.85$ safe floor.}
\label{fig:convergence}
\end{figure}

Figure~\ref{fig:convergence} shows how precisely the two ensembles
pin down that cosmic variance. We bootstrap-resample $N$
realizations, with replacement, from the 200 Patchy and the 25
Abacus ELG realizations and recompute the relative cosmic-variance
scatter $\sigma\,/\,\langle n(>R) \rangle$ at three representative
radii. The central value is essentially flat in $N$ and we omit it;
the top panel shows the relative uncertainty on the estimate. For
Patchy it falls from $\sim 35\%$ at $N = 5$ to $\sim 5\%$ for the
full 200-realization ensemble, tracking to within $\sim 10\%$ the
$1/\sqrt{2(N-1)}$ expected for the standard deviation of
Gaussian-distributed counts. The 25 Abacus ELG realizations
determine their cosmic-variance scatter to $\sim 10$--$13\%$.

The bottom panel of Figure~\ref{fig:convergence} translates this
into a practical mock-count budget at fixed data-vector length $K$,
via the Hartlap precision-matrix bias factor
$f = (N - K - 2)/(N - 1)$~\citep{Hartlap2007}. The
200 Patchy realizations supply $f \geq 0.85$ for $K \leq 28$,
comfortable for any single-statistic void analysis. Joint-probe
analyses concatenating multiple void summaries with galaxy
clustering or lensing push $K$ to several tens: reaching
$f = 0.85$ at $K = 60$ takes $N \approx 400$, and $N \sim 1000$ gives
$f \approx 0.94$. At BOSS scale, 1000 realizations take roughly a day
of laptop wall-time at the per-realization cost of
Section~\ref{sec:scaling}.

\begin{figure*}[tp]
    \plotone{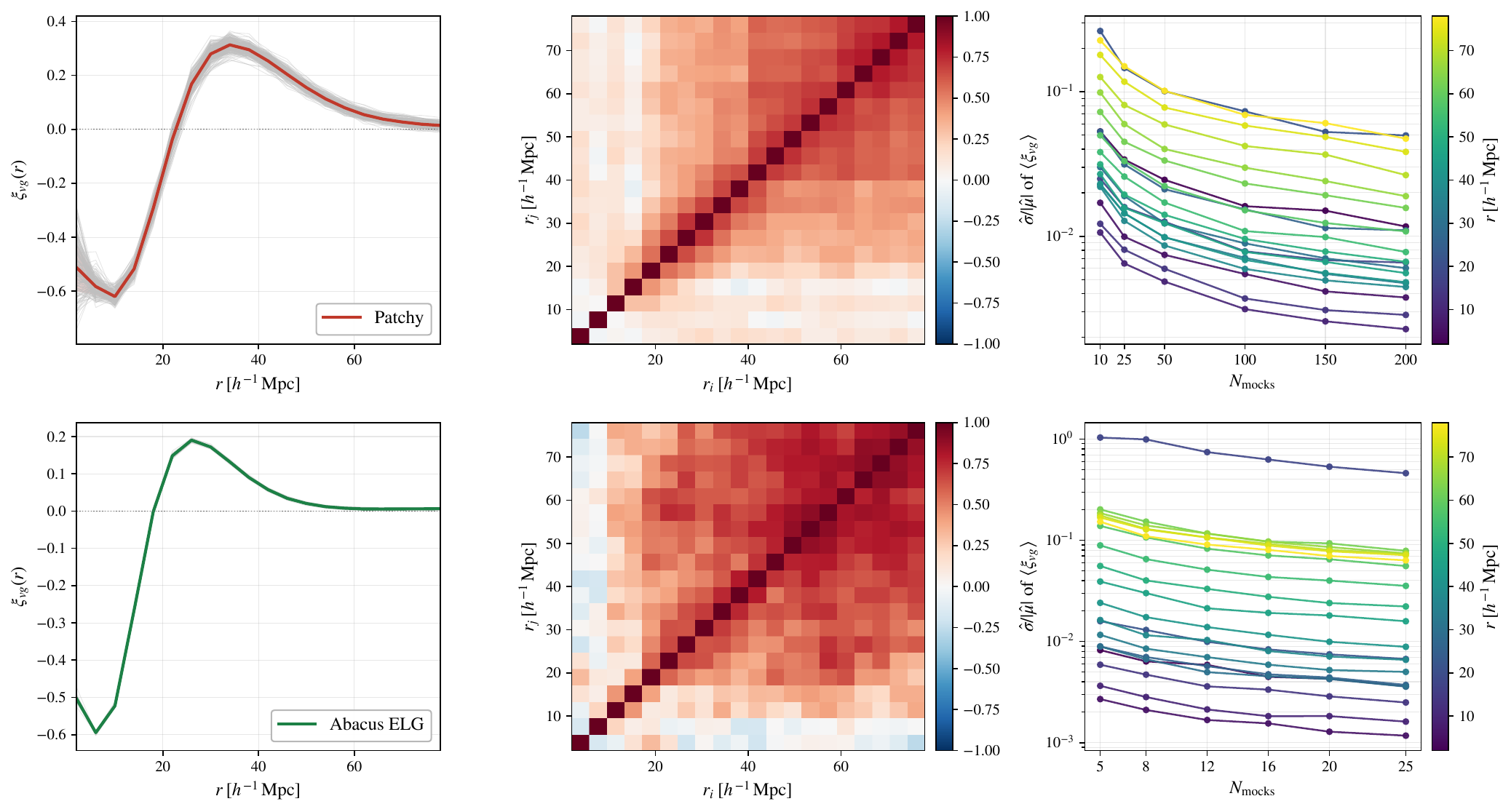}
  \caption{Void-galaxy monopole cross-correlation $\xi_{vg}(r)$ from
           both demonstration ensembles: Patchy (top, 200
           realizations) and Abacus ELG (bottom, 25 realizations).
           Both measurements use the central voids whose
           macrocenter lies at least $3R$ from any survey edge.
           \emph{Left:} per-mock
           curves (light grey) with the ensemble mean and
           $\pm 1\sigma$ cosmic-variance band. \emph{Middle:}
           correlation matrix $C_{ij}/\sqrt{C_{ii}C_{jj}}$ of the
           20-bin estimator. \emph{Right:} bootstrap estimate of
           $\hat\sigma/|\hat\mu|$ on the ensemble-mean estimator
           versus $N_\mathrm{mocks}$ ($N_\mathrm{boot} = 200$
           resamples per ensemble size), color-coded by separation
           $r$.}
\label{fig:xivg_cov}
\end{figure*}

The \ap~test remains a prime target for capturing cosmological parameter information from cosmic voids in Stage IV surveys. 
From the data side, this requires the stacked void-galaxy
cross-correlation measured along and across the line of sight,
$\xi_{vg}(r_\parallel, r_\perp)$, together with an ensemble-derived
estimate of its covariance. As a first step,
Figure~\ref{fig:xivg_cov} shows the monopole $\xi_{vg}(r)$ and its
covariance, computed on the central voids
of each ensemble that additionally satisfy a stricter shell-coverage
cut: the macrocenter must lie at least $3\,R_v$ from any survey
edge, so the analysis shell out to $r/R_v = 3$ remains inside the
survey for every stacked void. The left panels show the
per-realization curves with the ensemble mean and $\pm 1\sigma$
band; both ensembles show the familiar compensated void profile,
with its depth and ridge location set by each sample's tracer
density and redshift. The middle panels are the
correlation matrices $C_{ij}/\sqrt{C_{ii}C_{jj}}$ of the 20-bin
estimator. They show strong positive correlations: adjacent bins
correlate at $\approx 0.6$, and beyond $r \approx 25\,h^{-1}\,$Mpc
every pair of bins is positively correlated (mean $0.55$ for Patchy
and $0.69$ for Abacus ELG) --- expected from bin coupling when voids
are large compared to the bin width, rather than from contamination. The right panels show the
bootstrap relative uncertainty $\hat\sigma / |\hat\mu|$ on the
ensemble mean as a function of $N_\mathrm{mocks}$. With all 200
Patchy realizations, about half the bins fall below 1\% (median
0.8\%); the exceptions are the bin nearest the zero crossing, where
$|\hat\mu|$ approaches zero, and the largest separations, which
reach $\sim 5\%$. The 25 Abacus ELG realizations reach a median of
1.3\%, and their curves are still falling at $N = 25$. The Hartlap
factors
are $f \approx 0.89$ for Patchy ($K = 20$, $N = 200$) and $f
\approx 0.13$ for Abacus ELG ($K = 20$, $N = 25$); the former is
comfortably inside the validity regime, the latter satisfies the
threshold $N > K + 2$ but is small, and a downstream analysis on
the Abacus ELG ensemble would either restrict $K$ or extend the
ensemble further.

\begin{figure*}[tp]
    \plotone{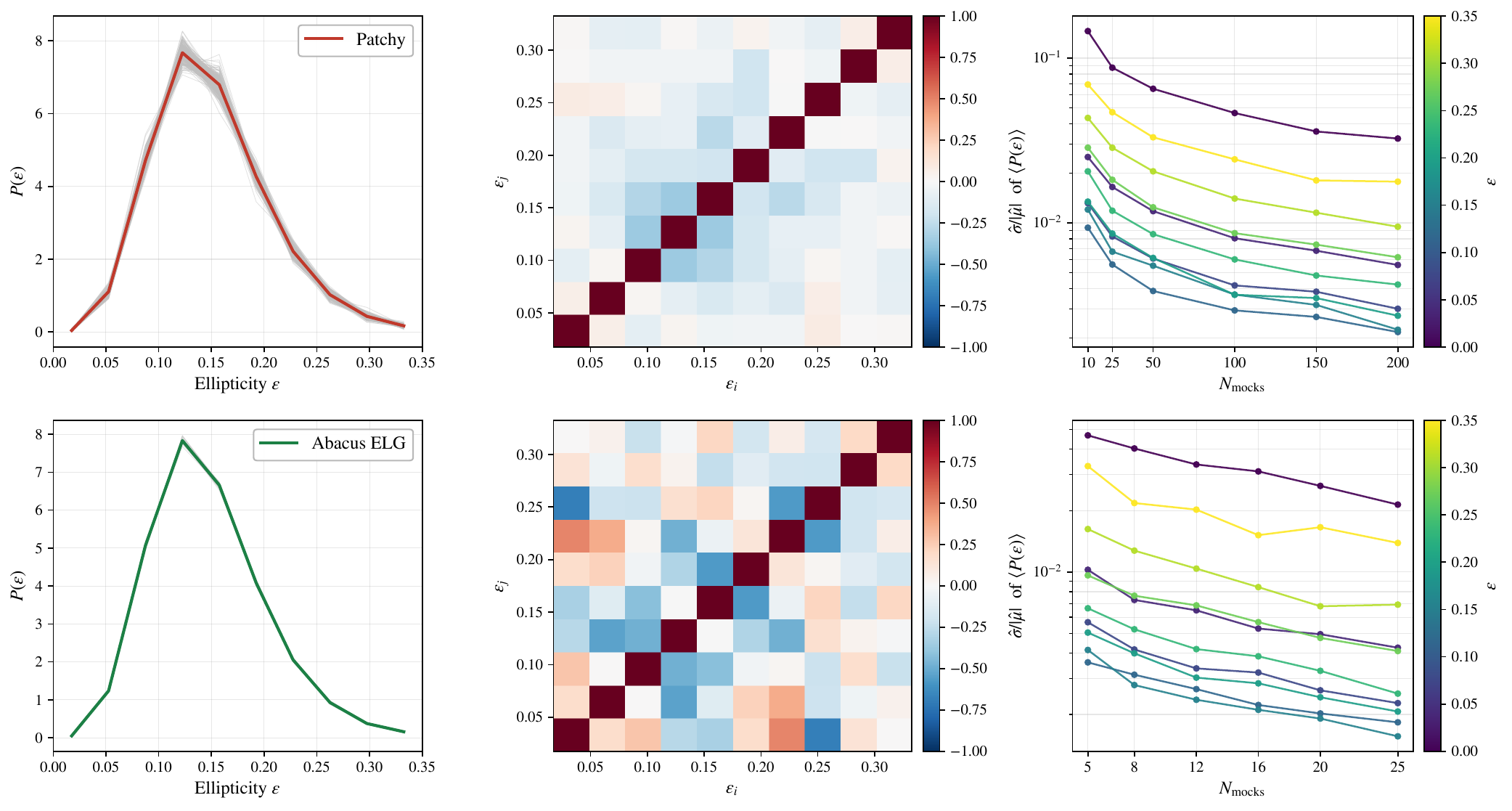}
  \caption{Ensemble void ellipticity distribution $P(\epsilon)$
           computed from each void's shape-tensor
           eigendecomposition, restricted to clean voids. Top:
           Patchy (200 realizations). Bottom: Abacus ELG (25
           realizations). Panels and
           color coding as in Figure~\ref{fig:xivg_cov}, on a 10-bin
           estimator.}
\label{fig:ellip_cov}
\end{figure*}

Watershed voids also carry shape information. As an example, we
apply the same analysis to the void ellipticity distribution
$P(\epsilon)$, computed from the shape-tensor eigendecomposition of
each clean void. Figure~\ref{fig:ellip_cov} shows the result, using
the same three-panel layout as Figure~\ref{fig:xivg_cov} but with
10 bins. Both ensembles peak near $\epsilon \approx 0.12$ with a
tail extending to $\epsilon \sim 0.30$. For Patchy the correlation
matrix is diagonal-dominant with mild anti-correlation between
adjacent bins (mean $-0.10$), as expected for a normalized histogram,
where an excess in one bin must be offset elsewhere; the Abacus ELG
matrix shows the same pattern but is noisier, as expected from only
25 realizations. We include $P(\epsilon)$ because
spherical-overdensity finders assume every void is a sphere, so they
cannot produce this statistic at all; we can measure it only because
the watershed approach defines each void by its actual basin
geometry.

\begin{figure}[htbp]
    \plotone{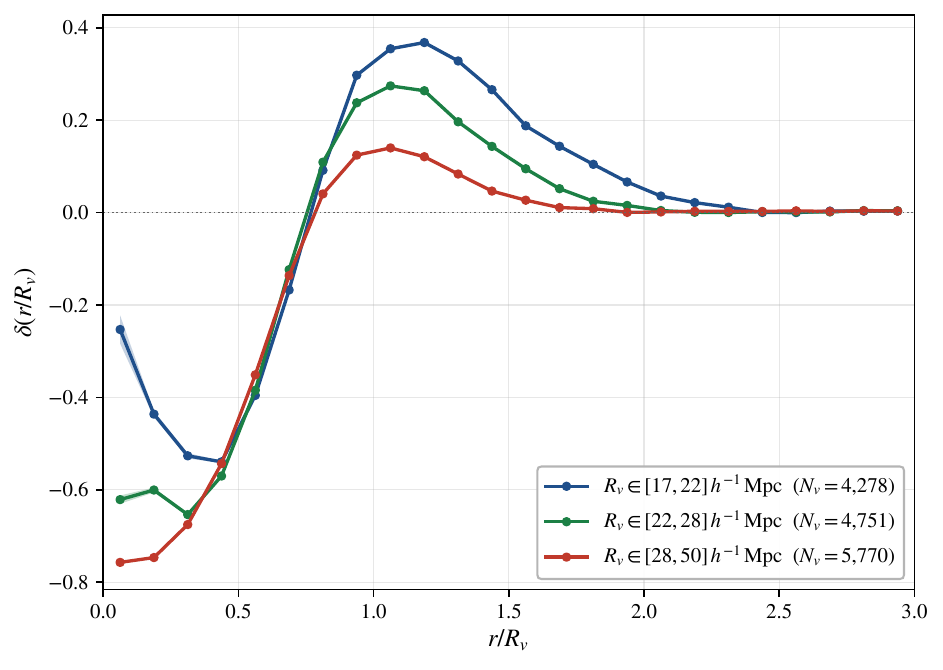}
  \caption{Stacked galaxy density profile $\delta(r/R_v)$ around the
           central voids of the Abacus ELG mock0 realization
           ($1.21 \times 10^{7}$ galaxies, $f_{\rm sky} = 0.22$,
           $z \in [0.8, 1.6]$) whose macrocenter lies at least
           $3\,R_v$ from any survey edge, in three bins of void
           radius. Bands are 10-fold spatial-jackknife errors,
           mostly smaller than the markers. Pair counts are computed
           against the Abacus ELG random catalog
           ($1.83 \times 10^{7}$ points) projected to comoving
           Cartesian coordinates with $\Omega_m = 0.315$.}
\label{fig:quad_flex}
\end{figure}

We can also examine what \videtwo~produces on Stage~IV-class
catalogs using only its default settings.
Figure~\ref{fig:quad_flex} shows the stacked galaxy density profile
$\delta(r/R_v)$ in three void-radius bins, around the central voids
of the Abacus ELG mock0 realization whose macrocenter lies at least
$3\,R_v$ from any survey edge. Pair counts are estimated against the
Abacus ELG random catalog ($1.83 \times 10^{7}$ points across NGC +
SGC), and errors come from a 10-fold spatial jackknife. The profiles
recover the universal void shape of
\citet{Hamaus2014}: an interior underdensity that
deepens with void size (minimum $\delta \approx -0.54$ to $-0.76$),
a zero crossing at $r/R_v \approx 0.75$--$0.78$, and a compensation
overdensity near $r/R_v \approx 1.1$ whose amplitude decreases
monotonically with void size, from $+0.37$ for
$R_v \in [17, 22]\,h^{-1}\,$Mpc to $+0.14$ for
$R_v \in [28, 50]\,h^{-1}\,$Mpc, before returning to the survey mean
by $r/R_v \approx 2$. The amplitude--$R_v$ relation is in
quantitative agreement with the BOSS~DR11 CMASS stacked-watershed
measurements of \citet{Hamaus2016}.

In absolute units, the ensemble-mean cross-correlation
$\xi_{vg}(r)$ over all 25 Abacus ELG realizations (bottom row of
Figure~\ref{fig:xivg_cov}; $17{,}857 \pm 126$ voids per realization
under the same cut) reaches $\xi_{vg} \approx -0.59$ at
$r \approx 6\,h^{-1}\,$Mpc, crosses zero near
$r \approx 18\,h^{-1}\,$Mpc, peaks in compensation at
$\xi_{vg} \approx +0.19$ near $r \approx 26\,h^{-1}\,$Mpc, and falls
below $0.01$ by $r \approx 60\,h^{-1}\,$Mpc. The scatter between
realizations is small: beyond the innermost bin,
$\sigma \lesssim 0.005$ everywhere, under $1\%$ of the trough depth
and about $2\%$ of the compensation peak.

\begin{figure}[htbp]
    \plotone{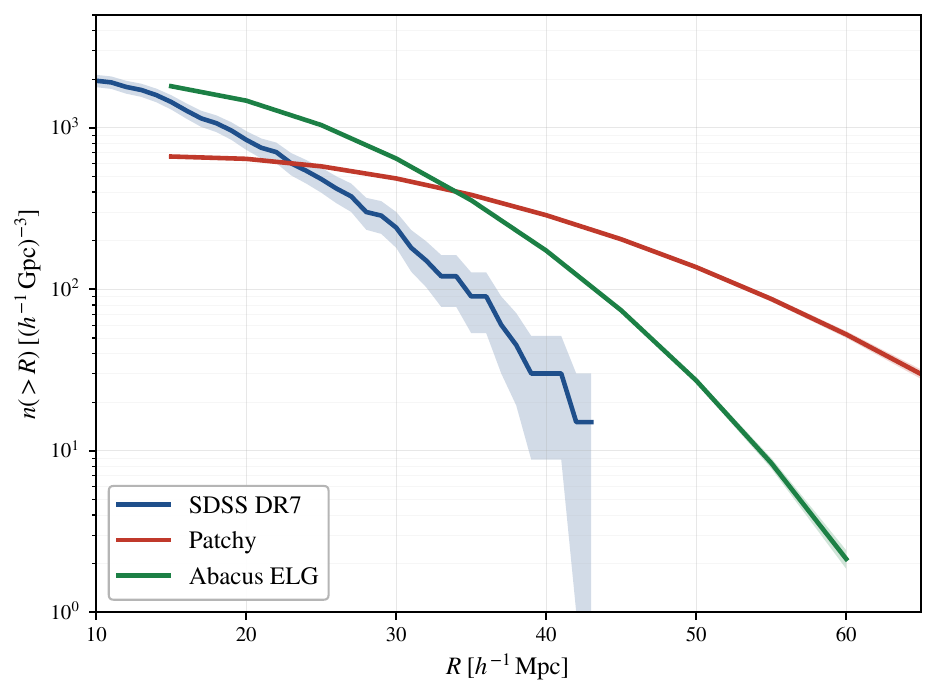}
  \caption{Cumulative void number function $n(>R)$ for clean voids
           across the three
           catalog products demonstrated in this work: a single
           observed SDSS~DR7 sample at the auto-resolved
           \nside~(blue), the 200-realization Patchy ensemble mean
           (red, with cosmic-variance $\pm 1\sigma$ band), and the
           25-realization Abacus ELG ensemble mean (green, same
           band). The two ensemble curves cross at $R \approx
           34\,h^{-1}\,$Mpc.}
\label{fig:unified_nfunc}
\end{figure}

Lastly, we can compare the three catalogs directly.
Figure~\ref{fig:unified_nfunc} overlays
the cumulative number function $n(>R)$ of clean voids on a single
axis for the observed SDSS~DR7 sample (a single catalog), the
Patchy ensemble mean, and the Abacus ELG ensemble mean.
Note that the two ensemble curves cross at $R \approx
34\,h^{-1}\,$Mpc: the Abacus ELG mocks produce $\sim 2$--$3\times$ more
small voids than Patchy at $R \lesssim 20\,h^{-1}\,$Mpc, but Patchy
extends to larger $R$ with $\sim 25\times$ more voids by
$R = 60\,h^{-1}\,$Mpc. Three effects drive this. The dominant one
is tracer density: watershed voids are built on the Voronoi
tessellation of the tracers, so the characteristic void size scales
with the mean tracer separation $\bar s = n_g^{-1/3}$. Patchy and
Abacus ELG differ in $\bar s$ by a factor of $\sim 1.4$
($\bar s_\mathrm{Patchy} \approx 17\,h^{-1}\,$Mpc,
$\bar s_\mathrm{Abacus} \approx 12\,h^{-1}\,$Mpc), so the sparser
Patchy catalog produces correspondingly larger Voronoi cells and
watershed basins. In each sample the smallest voids are a few times
the mean tracer separation, the standard \textsc{zobov} resolution
scaling~\citep{Neyrinck2008, Sutter2014}.
Two subdominant effects push in the same direction: cosmic
structure is further along at Patchy's median $z \approx 0.5$, where
the linear growth factor is $\sim 40\%$ larger than at the Abacus ELG
mocks' $z \approx 1.2$, and the
more strongly biased BOSS LRGs ($b \approx 2$) avoid void
interiors more cleanly than the less biased ELGs ($b \approx
1.3$--$1.5$). A controlled test would subsample the Abacus ELG
catalogs down to Patchy's tracer density and rerun;
we expect the small-$R$ surplus to disappear and the crossover to
migrate toward larger $R$.

A cosmological measurement is beyond the scope of this paper. A
defensible \ap~analysis on Stage~IV data
requires a theoretical template mapping cosmological parameters
onto $\xi_{vg}(r_\parallel, r_\perp)$ calibrated against
cosmology mocks, treatment of
void selection bias under the chosen tracer and cut
definition~\citep{Pisani2015, Hamaus_2017},
and marginalization over nuisance parameters within a
mock-validated likelihood. The contribution of this section
is to show that our deterministic void finder yields ensemble
covariances that converge, and that it delivers the multivariate
per-void inputs that void-based cosmological studies can adopt
without reprocessing.

\section{Conclusions}
\label{sec:conclusions}

We have presented \videtwo, an overhaul and upgrade to the \vide~void-finding and
analysis package. We primarily focused on improving and clarifying the
relationships among Voronoi tessellations, watershed algorithms, and the survey
mask. In addition to providing a rigorous procedure for mapping a
survey's boundary into the void finding algorithm, we have also significantly
improved the performance and reliability of the \vide~engine. In particular,
\videtwo~enjoys orders-of-magnitude speedups, significantly less memory usage,
and near-optimal scalability with respect to tracer number.

Void analysis will play a critical role in current and upcoming large galaxy
surveys, and with these upgrades in place \videtwo~is now poised to handle the
large quantities of data produced by these surveys. Additionally, \vide~is now
fully parameter-free, aside from user specification of the mask, allowing our
approach to identify voids in any survey configuration and
arrangement with maximum flexibility.

The approach of the VAST void-finding engine~\citep{Douglass2023} employs a
related strategy to survey edge-finding, wherein they flag any galaxy whose
local Voronoi polyhedron extends past the survey boundary, and then assign an
infinite density to those galaxies prior to the watershed step. However, to
determine which voids are contaminated by the edge, VAST allows the user to
specify what percentage of any void's surface area must include edge galaxies.
We adopt a simpler, parameter-free approach, wherein we mark any void that
includes even a single edge-flagged galaxy as \emph{contaminated}. However, any
voids that are merely adjacent to edge galaxies remain, since the interior
edges of Voronoi cells of those edge galaxies are secure.

The survey mask can have dramatic impacts on the derived void catalog, since voids
are large, extended objects that often intersect survey boundaries. A clean,
precise understanding of the mask, as viewed through the same morphological lens
as the watershed basins themselves, is crucial for properly navigating the
implications of a resulting void catalog. 

As discussed in Section~\ref{sec:masks_nside}, the choice of \nside~has
two failure modes that constrain it from above and below. We addressed
this by introducing an automatic rule that selects \nside~by matching
the mean tracer separation in a volume-weighted sense; for our
fiducial DR7 example the rule selects \nside~$= 64$.

The goal of \videtwo~is to allow the user maximum freedom to use the produced
void catalogs for their desired purposes. In the tradition established with the
release of \videone, the void finding package (housed at
\url{http://www.cosmicvoids.net}) includes many helpful Python-based analysis
tools and methods, which allow the user to quickly load void catalogs, member
galaxies, and more, as well as filter the catalogs to suit desired purposes and
perform analyses as they wish. Users can select from \emph{all} voids
(including ones that are contaminated by galaxies near the survey edge),
\emph{clean} voids free from edge effects, and \emph{central} voids
that are as bias-free as possible. Users may also choose to build an entire (sub-)void
hierarchy.

We have made available the visualization routines used
to create the new slice plots in this paper in the analysis package. All analysis
routines are fully backwards-compatible with void catalogs produced by
\videone.

In addition to the crucial boundary handling updates discussed in the present
paper, the public code
repository\footnote{\url{http://www.bitbucket.org/cosmicvoids}} details many
other updates, fixes, and quality-of-life improvements, such as more efficient
memory and storage handling, code reorganization and cleanup, a modern
interface for working with and manipulating void datasets, enhanced
capabilities for managing survey incompleteness and weighting schemes, and much
more. These details are discussed in the documentation made available as part of the public repository.

This is the first major upgrade to \vide~in over a decade. It maintains the
parameter-free spirit that first
animated the conversion of \vide~into a fully-featured void analysis suite.
Void catalogs produced by~\videtwo~are also fully deterministic:
given the same input galaxy catalog and survey mask, the output is
bit-for-bit identical across runs. This is in contrast to~\videone, where
stochastic boundary particle placement meant that void catalogs,
particularly near survey edges, varied between runs, complicating
reproducibility studies and covariance estimation from mock catalogs.
Furthermore,~\videtwo~stores per-galaxy edge provenance flags, enabling
downstream analyses to apply custom contamination criteria without
re-running the pipeline. Combined with the richer void property
estimates now computed during pruning (ellipticity, eigenvalues, density
profiles), the catalogs produced by~\videtwo~are not only more complete
but carry substantially more metadata per void, reducing the need for
post-hoc reprocessing.

Our analysis of two large mock suites --- 200 MultiDark-Patchy DR12~NGC
realizations at BOSS scale, and 25 AbacusSummit DESI~Y1 ELG
realizations at DESI~Y1 scale --- produced converged
cosmic-variance covariances for the void number function, the
void-galaxy cross-correlation $\xi_{vg}(r)$, and the void ellipticity
distribution $P(\epsilon)$. Because \videtwo~is deterministic,
none of the measured scatter comes from the void finder itself.

The full ensemble work was completed on commodity laptop hardware under modest process parallelism, with no
parameter retuning between the BOSS-class and DESI-class samples.
Together with the single-laptop scaling of Section~\ref{sec:scaling},
this makes Stage~IV void analyses, and the \ap~test in particular,
practical on modest computing resources.

Finally, \vide~maintains a long history of use, from a variety of
simulations to multiple surveys, providing the community with a rich
history of understanding as seen through the lens of this algorithm.
Follow-up papers will use the upgraded pipeline to make actual
cosmological measurements. Here we have shown that everything
leading up to those measurements, from deterministic catalog
production to ensemble-level covariance estimation, can be done on
commodity hardware.

\begin{acknowledgments}

PMS owes an enormous debt of gratitude to Guilhem Lavaux, Alice Pisani, Nico Hamaus, and all the other void enthusiasts who maintained \vide~over the years.
We thank Giovanni Verza for implementing per-tracer weights in \vide.
BDW acknowledges the support of the Simons Foundation through the Simons Collaboration on ``Learning the Universe''. 

\end{acknowledgments}

\software{\vide~\citep{Sutter2014},
          \textsc{voro++}~\citep{Rycroft2009},
          \healpix~\citep{Gorski2005_HEALPix},
          healpy~\citep{Zonca2019},
          NumPy~\citep{Harris2020},
          SciPy~\citep{Virtanen2020},
          Astropy~\citep{Astropy2013, Astropy2018, Astropy2022},
          Matplotlib~\citep{Hunter2007},
          h5py, HDF5, GSL, CFITSIO}

\bibliography{nuvide}
\bibliographystyle{aasjournalv7}

\end{document}